%% file: main.tex
\documentclass[twocolumn, journal]{IEEEtran}

\input{Preamble/preamble}

\begin{document}

\title{Optimal Topology Transition}

\author{Tong Han,\IEEEmembership{}
        David J. Hill,~\IEEEmembership{Life Fellow, IEEE,} and 
        Yue Song,~\IEEEmembership{Member, IEEE} 
\thanks{This work was supported in part by Hong Kong RGC General Research Fund under Project 17209419, and in part by the HKU Seed Fund for Basic Research for New Staff under Project 202009185007.}
\thanks{T. Han is with the Department of Electrical and Electronic Engineering, University of Hong Kong, Hong Kong and the School of Electrical and Electronic Engineering, Nanyang Technological University, Singapore 639798 (e-mail: hantong@eee.hku.hk).}
\thanks{Y. Song is with the Department of Electrical and Electronic Engineering, University of Hong Kong, Hong Kong (e-mail: yuesong@eee.hku.hk).}
\thanks{D. J. Hill is with the Department of Electrical and Computer Systems Engineering, Monash University, Melbourne, Australia and the Department of Electrical and Electronic Engineering, The University of Hong Kong (e-mail: david.hill3@monash.edu, dhill@eee.hku.hk).}
}

\maketitle

\begin{abstract}
  Network topology has significant impacts on operational performance of power systems. While extensive research efforts have been devoted to optimization of network topology for improving various system performances, the problem of how to transition from the initial topology to the desired optimal topology requires study. To address this problem, we propose the concept of optimal topology transition (OTT). This aims to find the topology transition trajectory from an initial topology to a desired terminal topology, which optimizes certain transition performance and satisfies operational constraints. The OTT problem is further formulated as a mixed-integer program under certain assumptions. Next, we propose the formulation of transition-embedded topology optimization that is capable of optimizing network topology and its transition trajectory simultaneously. Considering the time complexity of directly solving the mixed-integer programs, an efficient problem-specific solution algorithm is developed. Finally, numerical studies demonstrate the effectiveness of the proposed OTT and transition-embedded topology optimization models, as well as the superiority of the obtained optimal transition trajectories compared to ad hoc transition trajectories.
\end{abstract}


\IEEEpeerreviewmaketitle

\section*{Notation and Nomenclature}

$\!\!\!\!\!\!\!\!$
\textit{Notation}:$\!$ 
    \textit{(1)} For a vector $\bm{x}$, $\bm{x}\D$ is the diagonal matrix with
    entries of $\bm{x}$ on the main diagonal. For a square matrix $\bm{X}$, $\bm{X}\D$ is the vector of the main diagonal of $\bm{X}$. 
    \textit{(2)} $|\cdot|_{\circ}$ denotes the element-wise absolute value of a vector or matrix.
    \textit{(3)} For a set $\mathcal{X}$, $\vec{\mathcal{X}}$ is the vectorization of all elements in $\mathcal{X}$. 
    \textit{(4)} $\llbracket 1, n \rrbracket$ denotes the set of integers from 1 to $n$. 
    \textit{(5)} $\mathcal{X} \triangle \mathcal{Y}$ denotes the symmetric difference of sets $\mathcal{X}$ and $\mathcal{Y}$.
    \textit{(6)} For $x \!\in\! \mathbb{R}$, $I(x) \!=\! x$ if $x\!>\! 0$ and $0$ otherwise, and for $\bm{x} \!\in\! \mathbb{R}^n$, $I(\bm{x})$ is element-wise. 
    \textit{(7)} For $\bm{x} \!\in\! \mathbb{R}^n$, $\mathbb{V}(\bm{x}) \!\coloneqq\! \{ \bm{x}' \!\in\! \mathbb{R}^n | \bm{x}' \!\neq\! \bm{x}; \bm{x}'_i \!\neq\! 0 \!\Rightarrow\! \bm{x}'_i \!\neq\! \bm{x}_i  \} $.

\setlist[description]{labelindent=0pt,style=multiline,leftmargin=1.12cm}

\subsection{Sets and Graphs}
\addcontentsline{toc}{section}{Nomenclature} 
\begin{description} 
    \item[$\mathcal{E}_a, \mathcal{V}$] Set of edges (branches) and nodes (buses).
    \item[$\mathcal{E}_u, \mathcal{E}_s$] Set of unswitchable/switchable branches.
    \item[$\mathcal{E}_0, \mathcal{E}_T$] Set of branches that are switched on in the initial/ terminal topology. 
    \item[$\!\mathcal{G}(\mathcal{V}\!,\! \mathcal{E}_a)$] The undirected graph representing the power network topologically.
    \item[$\mathcal{P}$]  The set collecting all associated electrical properties.
    \item[$\mathcal{P}_p$] Set of electrical properties whose values are given in the topology optimization problem (TOP). 
    \item[$\mathcal{P}_{v}$] Set of unknown electrical properties in the TOP, i.e., $\mathcal{P}_{v} \!=\! \mathcal{P} \backslash \mathcal{P}_{p}$.
    \item[$\mathcal{P}_{vc}$] Set of electrical properties which are optimized coordinating with network topology in the TOP.   
    \item[$\mathbb{E}$] Feasible region of network topology.
    \item[$\mathbb{P}$] Feasible region of the unknown electrical properties.
    \item[$\mathbb{S}$] Feasible region of topology transition trajectories. 
\end{description}

\subsection{Scalars and functions}
\begin{description}
    \item[$\alpha_b, \alpha_v,\\ \alpha_{c}$] Positive weights for different components in the transition performance metric.
    \item[$\alpha _p, \alpha_n$] Positive weights for the penalty term and total number of line switching batches. 
    \item[$\beta$] Positive weight for the objective function of the TOP.  
    \item[$d_t$] The duration of topology $\mathcal{E}_t$.
    \item[$n_a$] Number of non-intersection subsets of switchable branches under the asynchronous switching mode.
    \item[$n_e$] The maximum allowable number of extra switching actions beyond the necessary switching. 
    \item[$n_o, n_{r, o}$] Total number of operational constraints that are allowed to be relaxed in $\mathbb{F}$ and $\mathbb{F}_r$, respectively.  
    \item[$n_p$] Number of unknown electrical properties in the TOP.
    \item[$n_s, n_s'$] Upper bounds of the number of lines being switched by OTS and the number of lines in the optimal topology of OTS.
    \item[$t_0, t_T$] Starting/end time of the during-transition stage. 
    \item[$F, f$] Objective functions in the TOP. 
    \item[$H(\cdot)$] Transition performance metric. 
    \item[$H_b, H_v$] Boundedness/volatility metric.
    \item[$H_c$] Transition cost.
    \item[$H_n$] Total number of line switching batches. 
    \item[$H_p$] Penalty term minimized together with the transition performance metric. 
    \item[$T, T_u$] The total number of transitional topologies plus 1 and its upper bound.
\end{description}

\subsection{Vectors and Matrices}
\begin{description} 
    \item[$\bm{\theta}$] $\!\in \mathbb{R}^{|\mathcal{V}|}$, phase angles of node voltages. 
    \item[$\overline{\bm{\theta}}_l, \overline{\bm{\theta}}_{lr}$]  $\!\in\!\! \mathbb{R}^{|\mathcal{E}_a|}$, absolute value limits of angle differences of branches for transitional/intermediate systems.
    \item[$\bm{\vartheta}, \bm{\rho}$] $\!\in\! \mathbb{R}^{|\mathcal{V}|}$, $\!\in\! \mathbb{R}^{|\mathcal{E}_a|}$, auxiliary variables in OTS formulation.  
    \item[$\bm{\vartheta}_t, \bm{\rho}_t$] $\!\in\!\! \mathbb{R}^{|\mathcal{V}|}$, $\!\in\! \mathbb{R}^{|\mathcal{E}_a|}$, auxiliary variables in OTT formulation. 
    \item[$\bm{b}_l, \bm{p}_{l}$] $\!\in\! \mathbb{R}^{|\mathcal{E}_a|}$, branch reactances and branch active powers.
    \item[$\bm{c}$] $\!\in\! \mathbb{R}^{|\mathcal{V}|}$, a constant uniquely-balanced vector. 
    \item[$\bm{p}_{d}, \bm{p}_{g}$] $\in \mathbb{R}^{|\mathcal{V}|}$, active powers of node loads/generations.
    \item[$\bm{p}_{g}^*$] $\in \mathbb{R}^{|\mathcal{V}|}$, the optimal value of active powers of generations given by the DC OTS model.
    \item[$\underline{\bm{p}}_{g}, \overline{\bm{p}}_{g}$] $\in \mathbb{R}^{|\mathcal{V}|}$, lower/upper bounds of active powers of node generations. 
    \item[$\overline{\bm{p}}_l, \overline{\bm{p}}_{lr}$] $\in\!\! \mathbb{R}^{|\mathcal{E}_a|}$, active power limits of branches for transitional/intermediate systems.
    \item[$\bm{w}_b,\! \bm{w}_v$] $\in \mathbb{R}^{n_p}$, vectors to scale different types of unknown electrical properties in the TOP and also to indicate importance of each property.
    \item[$\bm{w}_c$] $\!\in\! \mathbb{R}^{|\mathcal{E}_a|}$, line switching costs. 
    \item[$\vec{\mathcal{P}}_{v,t}, \\ \vec{\mathcal{P}}_{vr,t}$] $\in \mathbb{R}^{n_p}$, steady-state values of the unknown electrical properties in the TOP under topology $\mathcal{E}_t$ and any intermediate topology during the topology transition from $\mathcal{E}_{t-1}$ to $\mathcal{E}_{t}$. 
    \item[$\bm{\theta}_t, \\ \bm{\theta}_{r, t}$] $\!\in \mathbb{R}^{|\mathcal{V}|}$, steady-state values of phase angles of node voltages under topology $\mathcal{E}_t$ and $\mathcal{E}_{t-1} \!-\! \mathcal{E}_{t} \triangle \mathcal{E}_{t-1}$.
    \item[$\bm{p}_{l, t}, \\ \bm{p}_{lr, t}$] $\!\in\! \mathbb{R}^{|\mathcal{E}_a|}$, steady-state values of branch active powers under topology $\mathcal{E}_t$ and $\mathcal{E}_{t-\!1} \!-\! \mathcal{E}_{t} \triangle \mathcal{E}_{t-\!1}$.
 
    \item[$\bm{E}$] $\in\!\! \mathbb{R}^{|\mathcal{V}| \!\times\! |\mathcal{E}_a|} $, oriented incidence matrix of graph $\mathcal{G}(\mathcal{V}\!, \mathcal{E}_a)$ with each edge assigned an arbitrary orientation.  
\end{description}

\section{Introduction}

Network topology can significantly impact operational performance of power systems. In different voltage levels, these impacts have been leveraged to improve economy, security and stability of operations, which is commonly achieved by optimal transmission switching (OTS) or distribution network reconfiguration (DNR) \cite{4-62, 4-361, 4-811, 4-913}. They are essentially a \textit{topology optimization problem} (TOP): for given system parameters, forecasted variables and initial network topology, find a network topology that satisfies operational constraints to minimize (or maximize) an objective function, e.g., dispatch cost, power loss or stability level.

In the past decades, while research efforts have been devoted to the TOP extensively, focusing on application scenarios, formulations and solution approaches \cite{4-859, 4-61, 4-807}, one issue---\textit{how to transition from the initial topology to the optimal topology}---remains. 
Intuitively, the transition should not be arbitrary. For one thing, although the initial and optimal topologies both satisfy operational constraints, the intermediate topologies during the transition may not. For another, the transition process consists of line switching actions, which are essentially large-disturbances perhaps causing stability issues. However, to the best of our knowledge, this \textit{topology transition problem} has received little particular attention so far. A related work is using generation redispatch to reduce rotor shaft impacts induced by a single transmission loop closure \cite{4-1285}. 
The well-studied power system restoration problem contains a similar task that restores network topology from the post-outage state, which however, is over a much longer timescale and has different mechanisms compared with the topology transition problem \cite{4-1350, 4-1351, 4-1352}. 
There are several reasons behind this gap. 
Firstly, topology optimization is traditionally executed infrequently. For example, the execution cycle of DNR is at least several days and generally several weeks or even several months. This low execution frequency indicates a long time window with the feasible region of topology transition being large enough so that operators can easily select an ad hoc one. 
In addition, in view of the low-frequency execution as well as relatively short duration of topology transition, slight violations of operational constraints during topology transition can be allowable. 
Secondly, conventional power systems are commonly unstressed with regular power flow patterns, which diminishes the necessity of particular attention to topology transition. 
Thirdly, topology transition for DNR has little impacts on stability of conventional distribution networks with only passive loads. For OTS, electromechanical stability is usually unaffected by the transition process since the associated timescales can be separated if the transition is fast enough.

While low priority in the past, the topology transition problem becomes increasingly crucial with the transformation to power grids with high penetration of variable renewable energy (VRE). First of all, constant and dramatic changes in VRE promote more frequent and even dynamic real-time topology optimization to maximize profits \cite{4-913, 4-583, 4-905}. This complicates the selection of an ad hoc feasible transition and motivates optimization of the transition process. In addition, power flow patterns become highly diverse due to uncertainties in VRE together with changes in demand, and power grids are stressed both locally and globally. Accordingly, the feasible region of topology transition is potentially highly time-varying. 
On top of that, dynamic properties of inverter-dominant or -based power systems complicate topology transition. For transmission networks, inverter dynamics enlarge the timescale of structure-dependent stability phenomena to a period from several milliseconds to tens of seconds \cite{4-1302}. Hence, even fast transitions can induce instability. In medium/low voltage networks, inverter-based distributed generators also introduce new dynamics complicating the entire system dynamics \cite{4-990}. For example, in droop-based microgrids, it is observed that fast line dynamics also have a major influence on stability of slower modes \cite{4-891}, and network loops can reduce the stability margin \cite{4-620}. These structure-related observations contrast with our conventional view, which calls for at least a recheck of stability issues of the topology transition and probably a special topology transition controller to ensure stability. Finally, for the distributed topology optimization being developed recent years \cite{4-783, 4-822}, fast topology transition is far from easy to achieve due to the lack of a control center. Thus, a topology transition process necessarily consists of a sequence of line switchings with relatively long time intervals. 


In light of the indispensability of particular attention to the topology transition problem with power grid transformation, this paper proposes the concept of \textit{optimal topology transition} (OTT), aiming to find the topology transition trajectory which satisfies operational constraints and optimizes certain transition performance. 
Although a comprehensive framework for OTT should take both static and dynamic factors into consideration, this work, as the first step towards the goal, is concerned with a primary version that only addresses the static performance during transition. 
Firstly, some basics related to topology transition are defined and investigated, mainly including TOPs from which the topology transition problem emerges, switching modes, line switching batches, the general process and basic requirements of topology transition. 
In addition, ad hoc topology transition trajectories are analysed. 
Next, we define the problem of OTT, which is further mathematically formulated as a mixed-integer program with a focus on during-transition performance. We also develope a formulation of transition-embedded topology optimization in order to obtain the optimal topology and its transition trajectory simultaneously. To efficiently solve the mixed-integer programs, a problem-specific solution algorithm is designed. 
Finally, effectiveness of the proposed OTT and transition-embedded topology optimization models are demonstrated numerically in the context of various TOPs. It is found that ad hoc topology transition can cause violations of operational constraints and increases of dispatch cost, which can be eliminated completely and relieved by the obtained optimal transition trajectories, respectively.
 
The rest of the paper is organized as follows. Section II introduces some basics related to OTT and investigates possible ad hoc transition trajectories. Section III defines the problem of OTT and derives its mathematical formulation, and Section IV focuses on transition-embedded topology optimization and efficient solution algorithms. Numerical results are provided in Section V followed by Section VI making a conclusion and a prospect for future works.




\section{Topology Optimization Problem}

We first consider the TOP from which the topology transition problem emerges. It is noted that TOP here is used as a collective name to refer to OTS problems in transmission networks and DNR problems in distribution networks \cite{4-62, 4-361, 4-811, 4-913}. A more formal definition and a uniform formulation for TOPs, which tend to cover most cases and capture the most essential compositions of TOPs, are introduced in the following.

\begin{definition}[TOP]\label{def-6-1-1}
  $\!$Given an initial topology $\mathcal{E}_{0}$, values of $\mathcal{P}_{p} $, feasible regions $\mathbb{E}$ and $\mathbb{P}$, 
  find the topology $\mathcal{E} \!\!\subseteq\!\! \mathcal{E}_{\!a}$ satisfying $\mathcal{E} \!\!\in\!\! \mathbb{E}(\mathcal{E}_0)$ such that graph $\mathcal{G}(\mathcal{V}, \mathcal{E})$ minimizes $F(\mathcal{E} | \mathcal{E}_0, \mathcal{P}_p) \!\!=\!\! \min_{ \mathcal{P}_v  \in \mathbb{P}(\mathcal{E}, \mathcal{P}_p) } f(\mathcal{P}_v | \mathcal{E}, \mathcal{E}_0, \mathcal{P}_p)  $. 
  Alternatively, by introducing vector $\bm{z} \!=\! \col\{ \bm{z}_e \!\in\! \mathbb{B} |_{e \in \mathcal{E}_a } \}$ to sign elements of $\mathcal{E}$, i.e., $\bm{z}_e = 1$ if $e \in \mathcal{E}$ and $\bm{z}_e \!=\! 0$ otherwise, and letting $\bm{z} \!=\! \bm{z}_0$ for $\mathcal{E} \!=\! \mathcal{E}_0$\footnote{We will use $\bm{z}$ and $\mathcal{E}$ interchangeably referring to the same topology, which with the same subscript and superscript also refer to the same.}, the TOP is to find
\begin{subequations}\label{eq-6-1-1} 
  \begin{align}
    \bm{z}_T =&  \argmin\nolimits_{\bm{z}}~  F ( \bm{z} | \bm{z}_0, \mathcal{P}_p ) ~ \rm{s.t.}: \label{eq-6-1-1:1}\\
    & \bm{z} \in \mathbb{E}( \bm{z}_0 )  \label{eq-6-1-1:2}\\ 
    & \begin{aligned}
      & F ( \bm{z} | \bm{z}_0, \mathcal{P}_p )  =  \min\nolimits_{\mathcal{P}_v \in \mathbb{P}(\bm{z}, \mathcal{P}_p)  } f(\mathcal{P}_v | \bm{z}, \bm{z}_0, \mathcal{P}_p ) 
    \end{aligned} \label{eq-6-1-1:4}
  \end{align}
\end{subequations}  
\end{definition}

\begin{remark}
  \text{(\romannumeral1)} For most TOPs, $\mathbb{E}$ should be shaped at least by $\mathcal{E} \supset \mathcal{E}_u $, and may also by boundedness of $|\mathcal{E} \triangle \mathcal{E}_0|$. For OTS, $\mathbb{E}$ generally is also shaped by connectedness of $\mathcal{G}(\mathcal{V}, \mathcal{E})$ and boundedness of $|\mathcal{E}|$. For DNR, $\mathbb{E}$ is also shaped by forestness or treeness of $\mathcal{G}(\mathcal{V}, \mathcal{E})$ and pathless between any pair of substation nodes in $\mathcal{G}(\mathcal{V}, \mathcal{E})$. 
  \text{(\romannumeral2)} Domain $\mathbb{P}$ is determined by power flow and operational constraints.
  \text{(\romannumeral3)} Specific elements of $\mathcal{P}_p$ and $\mathcal{P}_v$ are problem-dependent. Network topology can be optimized independently or coordinating with properties in $\mathcal{P}_{vc}$ (call them coordinated properties). For example, for the OTS problem where generator outputs and network topology are optimized simultaneously, $\mathcal{P}_{vc}$ refers to power injections at generator nodes.
\end{remark}

To make the TOP and later formulations more comprehensible, a specific TOP version is provided. 
According to the network properties of different voltage levels, the TOP (\ref{eq-6-1-1}) can be generally simplified as a mixed-integer convex program (MICP). For example, for transmission networks, a typical TOP version, called DC OTS, employs the DC power flow model and augments economic dispatch with optimization of network topology. 
For DC OTS, we have 
\begin{equation*}\!\!
    \mathcal{P}      \!=\!\! \{ \bm{\theta}, \bm{p}_g, \bm{p}_d, \bm{b}_l\}, 
    \mathcal{P}_{\!p}    \!=\!\! \{ \bm{p}_{d}, \bm{b}_l \}, 
    \mathcal{P}_{\!v}    \!=\!\! \{ \bm{p}_{g}, \bm{p}_{l}, \theta \}, 
    \mathcal{P}_{\!vc} \!=\!\! \{ \bm{p}_{g} \},
\end{equation*}
and (\ref{eq-6-1-1}) is specified as 
\begin{subequations}\label{eq-6-1-1-add} 
    \begin{align}
      \bm{z}_T =&  \argmin\nolimits_{\bm{z} }~  F ( \bm{z} | \bm{z}_0, \bm{p}_{d}, \bm{b}_l ) ~ \rm{s.t.}: \label{eq-6-1-1-add:1}\\
      & \bm{E}_u \bm{z} = \bm{1}, \Vert \bm{z} - \bm{z}_0 \Vert_1 \leq n_s, \Vert \bm{z} \Vert_1 \geq n_s'    \label{eq-6-1-1-add:2} \\  
      & - M (\bm{1} - \bm{z}) \leq \bm{E}^T \bm{\vartheta} - \bm{\rho} \leq M (\bm{1} - \bm{z}) \label{eq-6-1-1-add:3} \\
      & - M \bm{z} \leq \bm{\rho} \!\leq\! M \bm{z}, \bm{E} \bm{\rho} \!=\! \bm{c},  \bm{\vartheta}\!\in\! \mathbb{R}^{|\mathcal{V}|},  \bm{\rho} \!\in\! \mathbb{R}^{|\mathcal{E}_a|} \label{eq-6-1-1-add:4} \\
      & F ( \bm{z} | \bm{z}_0, \bm{p}_{d}, \bm{b}_l )  =  \min\nolimits_{ \bm{\bm{p}_{g}, \bm{p}_{l}, \theta} } f( \bm{p}_g ) ~ \rm{s.t.}: \label{eq-6-1-1-add:5} \\
      &~~~~~~~~~~ - M (\bm{1} \!-\! \bm{z}) \!\leq\! \bm{b}_l\D \bm{E}^T \bm{\theta}  \!-\! \bm{p}_l  \!\leq\! M (\bm{1} \!-\! \bm{z}) \label{eq-6-1-1-add:6} \\
      &~~~~~~~~~~ \bm{E} \bm{p}_l = \bm{p}_g - \bm{p}_d \label{eq-6-1-1-add:7} \\
      &~~~~~~~~~~ M(\bm{z} \!-\! \bm{1}) \!-\! \overline{\bm{\theta}}_{l} \!\leq\! \bm{E}^T \bm{\theta} \!\leq\! \overline{\bm{\theta}}_{l} \!+\!  M( \bm{1} \!-\! \bm{z}) \label{eq-6-1-1-add:8} \\
      &~~~~~~~~~~  - \bm{z}\D~\! \overline{\bm{p}}_{l} \leq \bm{p}_l \leq  \bm{z}\D~\! \overline{\bm{p}}_{l} \label{eq-6-1-1-add:9} \\
      &~~~~~~~~~~ \underline{\bm{p}}_g \leq \bm{p}_g \leq \overline{\bm{p}}_g  \label{eq-6-1-1-add:10}
    \end{align}
\end{subequations}  
where (\ref{eq-6-1-1-add:2})-(\ref{eq-6-1-1-add:4}) form the feasible region of $\bm{z}$, i.e., $\mathbb{E}$, with (\ref{eq-6-1-1-add:3})-(\ref{eq-6-1-1-add:4}) being network connectedness constraints \cite{4-995-ea}; (\ref{eq-6-1-1-add:6})-(\ref{eq-6-1-1-add:10}) form the feasible region $\mathbb{P}$, with (\ref{eq-6-1-1-add:6})-(\ref{eq-6-1-1-add:7}) being DC power flow constraints, (\ref{eq-6-1-1-add:8})-(\ref{eq-6-1-1-add:10}) operational constraints for phase angle differences, active power of branches and active power of generations, respectively; and $f( \bm{p}_g )$ denotes the total generation cost, which is generally a piece-wise linear or quadratic function. 
Analogously, adopting the linear-programming approximation (LPA) based power flow model of transmission networks yields the LPA-based OTS where voltage magnitudes and reactive power are considered \cite{4-1163, 4-1129}. For distribution networks, the DistFlow model can be adopted to derive the DistFlow-based DNR \cite{4-811}.

\section{Basics of Topology Transition and Ad Hoc Transition Trajectories}

This section establishes some basics of topology transition and gives some possible ad hoc topology transition trajectories. 

\subsection{Basics of Topology Transition}

Two line switching modes and some concepts related to topology transition are first introduced.

\begin{definition}[Synchronous switching (SS) mode]
  $\forall \mathcal{E} \subseteq \mathcal{E}_s$, lines in $\mathcal{E}$ can be switched synchronously.
\end{definition}

\begin{definition}[Asynchronous switching (AS) mode]
  With switchable line set $\mathcal{E}_s$ divided into $n_a$ non-intersecting subsets $\{\mathcal{E}^i|_{i=1}^{n_{a}}\}$, $\forall \mathcal{E} \subseteq \mathcal{E}^i$, lines in $\mathcal{E}$ can be switched synchronously; and $\forall \mathcal{E} \subseteq \mathcal{E}^i $, $\mathcal{E}' \subseteq \mathcal{E}^j$, $i \neq j$, lines in $\mathcal{E}$ and that in $\mathcal{E}'$ can only be switched asynchronously.
\end{definition}

\begin{remark}
  \text{(\romannumeral1)} These two modes correspond to the centralized scenario where a control center executes topology transition, and the distributed scenario where each agent controls a set of lines $\mathcal{E}^i$, respectively. 
  \text{(\romannumeral2)} Here ``synchronous" and ``asynchronous" are distinguished as per the control center or agents send line switching commands synchronously or asynchronously instead of the actual line switching sequence. 
\end{remark}

\begin{definition}[Line switching batch, intermediate topology and transitional topology]
  The line switching batch refers to a group of synchronous line switching without any other line switching being synchronous with them. An intermediate topology refers to the topology resulting from any single line switching in the transition. Then for a given topology transition from $\mathcal{E}_0$ to $\mathcal{E}_T$, a transitional topology $\mathcal{E}_t$ with $t \!\in\! \llbracket 1, T\!-\!1 \rrbracket$, refers to the intermediate topology during the transition after any line switching batch.
\end{definition}

\begin{remark}
  A topology transition trajectory is fully determined by all its transitional topologies. Adding the known initial and terminal topologies, a topology transition trajectory can be represented by the topology sequence $(\mathcal{E}_0, \mathcal{E}_1,..., \mathcal{E}_{T - 1}, \mathcal{E}_T )$.
\end{remark}

\begin{figure}[h] 
	\centering
	\includegraphics[width = \columnwidth]{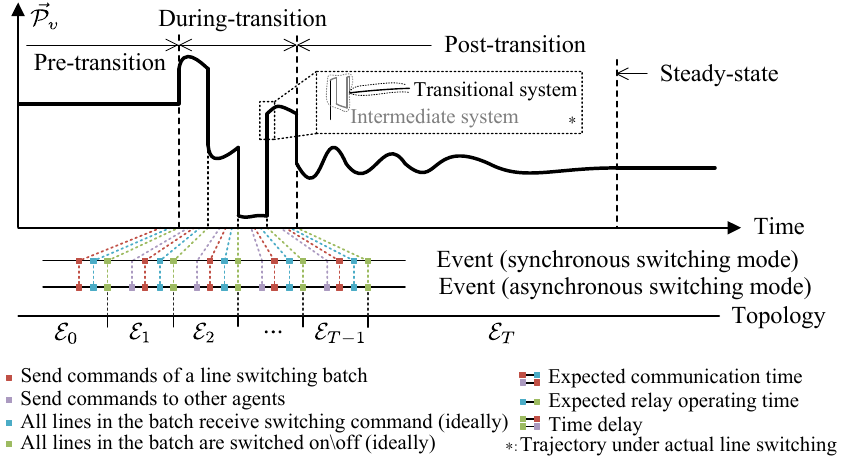}
	\caption{Illustration of the process of topology transition.}
  \label{fig-6-1-0}
\end{figure}

Fig. \ref{fig-6-1-0} illustrates the process of topology transition. For each line switching batch, the control center or agent firstly sends switching commands to all remote control units (RTUs) of lines in the batch. When RTUs receive switching commands, the corresponding lines will be switched on/off after a certain time interval determined by the relay operating time \cite{4-845}. Then, after a pre-set time delay, the control center or agent sends switching commands of next line switching batch. In particular, for the AS mode, if the next line switching batch is executed by the other agents, the agent finishing the current line switching batch will send commands to the other agents also after a pre-set time delay so that one of them can start executing the next line switching batch. It is noted that Fig. \ref{fig-6-1-0} corresponds to the ideal case with an identical communication time and an identical relay operating time, and thus all lines in any line switching batch are switched synchronously. However, the actual trajectory of $\vec{\mathcal{P}_v}$ would behave like that in the dotted frame due to uncertainties in communication time and relay operating time.

Additionally, Fig. \ref{fig-6-1-0} also shows fours stages related to topology transition: pre-transition, during-transition, post-transition and steady-state stages. In the pre-transition stage, the system is in a steady state determined by the initial topology $\mathcal{E}_0$. Then, the system state jumps after each line switching batch during topology transition. After execution of all line switching batches, the system state oscillates, and eventually converges to the steady state determined by the terminal topology $\mathcal{E}_T$ if its region of attraction contains the system state just before executing the last line switching batch, or diverges otherwise. 

It is assumed that in the pre-transition stage, $\mathcal{P}_{vc}$, if nonempty, has been tuned to its optimal value obtained by solving the TOP (\ref{eq-6-1-1}). Thus $\mathcal{P}_{p}$ and $\mathcal{P}_{vc}$ are both fixed in the during- and post-transition stages, and then we only need to consider the case where $\mathcal{P}_{vc} = \emptyset$. It is also assumed that the system in both pre-transition and steady-state stages satisfies normal operational constraints, namely that (\ref{eq-6-1-1:4}) is feasible with $\bm{z} = \bm{z}_0$ and $\bm{z} = \bm{z}_T$. Then topology transition should basically satisfy the following two requirements:
\begin{itemize} 
  \item[\textit{(\romannumeral1)}] All intermediate topologies are connected.
  \item[\textit{(\romannumeral2)}] In the during-transition stage, the system satisfies operational constraints.
\end{itemize}


Requirement \textit{(\romannumeral1)} comes from the topological requirements in TOPs. For OTS , connectedness of the terminal topology $\bm{z}_T$ is practically required since reconnecting and synchronizing islands of transmission networks could be very difficult and result in equipment failure \cite{4-761, 4-1283}. Thus, all intermediate topologies should be connected to prevent causing islands during topology transition. For DNR, given the radial terminal topology $\bm{z}_T$, islands during topology transition should also be prevented considering the increasing inverter dynamics in distribution networks. These dynamics can cause synchronization difficulties when reconnecting the main grid and an island and reconnecting different islands. 
Furthermore, for requirement \textit{(\romannumeral1)}, the actual sequence of line switching within a line switching batch is stochastic and thus intermediate topologies during two adjacent topologies from $(\mathcal{E}_0, \mathcal{E}_1,..., \mathcal{E}_{T - 1}, \mathcal{E}_T )$ are also uncertain. By considering all possible intermediate topologies, requirement \textit{(\romannumeral1)} is equivalent to the condition as follows:
\begin{condition}\label{cond-6-1-1}
  $\!\!\forall t \!\!\in\!\! \llbracket 1,\! T \rrbracket$, topology $\mathcal{E}_{t-\!1} \!\backslash\! (\mathcal{E}_{t} \triangle \mathcal{E}_{t-\!1})$ is connected.
\end{condition}

Requirement \textit{(\romannumeral2)} is necessary to ensure system operational security during topology transition. For this requirement, we first divide the power system in the during-transition stage into two kinds: the transitional system and intermediate system that under transitional topologies and other intermediate topologies, respectively, as illustrate in Fig. \ref{fig-6-1-0}.   
If topology transition is executed infrequently, some of the operational constraints here can be less stringent than that in (\ref{eq-6-1-1:4}). For example, the MVA rating of lines can be set to the emergency rating but not necessarily the long-term or short-term rating used by TOPs. However, for more frequent topology transition caused by frequent topology optimization, the operational constraints in requirement \textit{(\romannumeral2)} should be the same as or very close to normal operational constraints in (\ref{eq-6-1-1:4}). Moreover, since duration of the intermediate system is much shorter than that of the transitional system, it is reasonable to constrain the transitional system by normal operational constraints, i.e., those in (\ref{eq-6-1-1:4}); and the intermediate systems by operational constraints for emergency states, i.e., those in relaxed (\ref{eq-6-1-1:4}) defined by Definition \ref*{def-6-1-r1c}. 
\begin{definition}[Relaxed (\ref{eq-6-1-1:4})]\label{def-6-1-r1c}
    (\ref{eq-6-1-1:4}) with the bound parameters, e.g., $\overline{\bm{\theta}}_{l}$ and $\overline{\bm{p}}_{l}$ in (\ref{eq-6-1-1-add}), replaced by their counterparts for emergency states.
\end{definition}
Additionally, for variables in $\mathcal{P}_v$ concerned by operational constraints, including branch power and voltage magnitude, it is assumed that their during-transition trajectories under each topology can be approximated by their steady-state values under the corresponding topology. Hence, requirement \textit{(\romannumeral2)} is declared by Condition \ref{cond-6-1-2} and Condition \ref{cond-6-1-2-x}. 
 
\begin{condition}\label{cond-6-1-2}
  $\forall t \!\in\! \llbracket 1, T-1 \rrbracket$, (\ref{eq-6-1-1:4}) with $\bm{z} \!=\! \bm{z}_t$ is feasible.
\end{condition}

\begin{condition}\label{cond-6-1-2-x}
  $\forall t \!\in\!  \llbracket 1, T \rrbracket$ and $\Delta \bm{z}_t \in \mathbb{V}(\bm{z}_{t} - \bm{z}_{t-1})$, relaxed (\ref{eq-6-1-1:4}) with $\bm{z} \!=\! \bm{z}_{t-1} + \Delta \bm{z}_t$ is feasible.
\end{condition}

\subsection{Ad Hoc Topology Transition}

We use the term \textit{ad hoc} to indicate the topology transition trajectory \textit{promisingly} satisfies Condition \ref{cond-6-1-1} to Condition \ref{cond-6-1-2-x} without any calculation or with only simple calculation to check. Seeing that topologies $\mathcal{E}_0$ and $\mathcal{E}_T$ are both connected and make (\ref{eq-6-1-1:4}) feasible, according to Presumption \ref{asp-6-1-1} that indicates switching$\!$ on lines does not damage operational feasibility,$\!$ the simplest ad hoc topology transition under the SS mode is
\begin{equation}\label{eq-6-1-4}
    S_{syn} \triangleq (\mathcal{E}_0, \mathcal{E}_0 \cup (\mathcal{E}_T \backslash \mathcal{E}_0), \mathcal{E}_T)
\end{equation} 
which closes all lines in $\mathcal{E}_T \backslash \mathcal{E}_0$ first and then opens all lines in $\mathcal{E}_0 \backslash \mathcal{E}_T$, both by one line switching batch; and the simplest ad hoc topology transition under the AS mode is
\begin{equation}\label{eq-6-1-5}
  S_{asy} \!\triangleq\! (\mathcal{E}_0, \mathcal{T}_1, ..., \mathcal{T}_i, ..., \mathcal{E}_0 \cup (\mathcal{E}_T \backslash \mathcal{E}_0), \mathcal{T}'_1, ..., \mathcal{T}'_i, ..., \mathcal{E}_T)
\end{equation}
where all involved agents first sequentially close the respective controlled lines in $\mathcal{E}_T \backslash \mathcal{E}_0$ by one line switching batch, and then sequentially open the respective controlled lines in $\mathcal{E}_0 \backslash \mathcal{E}_T$ by one line switching batch. More formally, we have $\mathcal{T}_i \!=\! \mathcal{E}_0 \!\cup\! [ (\mathcal{E}_T \backslash \mathcal{E}_0) \!\cap\! \mathcal{A}_i ]$ and $\mathcal{T}'_i \!=\! (\mathcal{E}_0 \!\cup\! (\mathcal{E}_T \backslash \mathcal{E}_0) ) \backslash \mathcal{A}'_i $, where $\mathcal{A}_1 \!=\! \mathcal{E}^{j_1}$, $\mathcal{A}_i \backslash \mathcal{A}_{i - 1} \!=\! \mathcal{E}^{j_i}$ with $j_i \!\neq\! j_{i'}$ if $i \!\neq\! i'$, $j_i \!\in\! \llbracket 1, n_a  \rrbracket$ and $\mathcal{E}^{j_i} \!\cap\! (\mathcal{E}_T \backslash \mathcal{E}_0) \!\neq\! \emptyset$, and $\mathcal{A}'_i$ is similar to $\mathcal{A}_i$ but $\mathcal{E}^{j_i} \!\cap\! (\mathcal{E}_0 \backslash \mathcal{E}_T) \!\neq\! \emptyset$. 
By further splitting the line switching batch between two transitional topologies, other ad hoc topology transition trajectories can be obtained.

\begin{presumption}\label{asp-6-1-1}(Presumption for ad hoc topology transition)
    If (\ref{eq-6-1-1:4}) with topology $\mathcal{E}_0$ and $\mathcal{E}_T$ are both feasible, $\mathcal{E} \subseteq \mathcal{E}_T \backslash \mathcal{E}_0$ and $\mathcal{E}' \subseteq \mathcal{E}_0 \backslash \mathcal{E}_T$, then (\ref{eq-6-1-1:4}) with topology $\mathcal{E}_0 \cup \mathcal{E}$ and that with $\mathcal{E}_T \cup \mathcal{E}'$ are also feasible. 
  \end{presumption}

\begin{figure}[h]
	\centering
	\includegraphics[width=0.8\columnwidth]{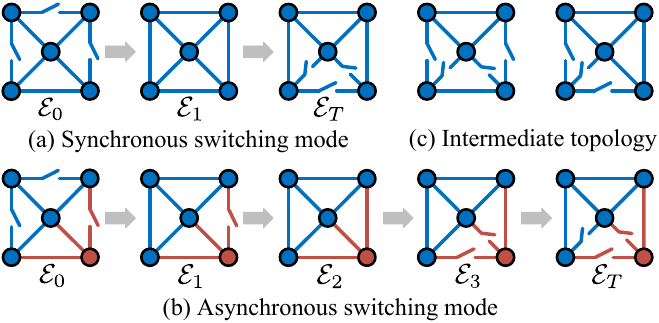}
  \caption{Illustration of ad hoc topology transition under (a) SS and (b) AS modes for a 5-bus system, and (c) possible intermediate topologies during an infeasible topology transition. For the AS mode, $\mathcal{E}_s$ is divided into $\mathcal{E}^1$ (blue lines) and $\mathcal{E}^2$ (red lines).}
  \label{fig-6-1-1}
\end{figure}

Fig. \ref{fig-6-1-1} shows implementations of the ad hoc topology transition to a 5-bus system under the two switching modes. Instead of the above ad hoc transition, the simplest transition is that realized by only one line switching batch $\mathcal{E}_0 \triangle \mathcal{E}_{T}$. Two possible intermediate topologies for this transition are given in Fig. \ref{fig-6-1-1} (c), which are  unconnected  due to the stochastic actual switching sequence of lines in $\mathcal{E}_0 \triangle \mathcal{E}_{T}$.

Another kind of ad hoc transition is based on the setting that the size of each line switching batch is 1, which corresponds to the legacy operational rule that actions of each line switching are executed sequentially. Although modern power grids with wide-area control or remote control are unrestricted to this rule, the search space of topology transition following it is relatively small. This kind of ad hoc topology transition is
\begin{equation}
  S_{\!one} \!\!\triangleq\!\! (\mathcal{E}_0, \mathcal{E}_1,..., \mathcal{E}_t, ..., \mathcal{E}_T)\!: T \!\!=\!\! |\mathcal{E}_0 \triangle \mathcal{E}_T| \land |\mathcal{E}_t \triangle \mathcal{E}_{t-1}| \!\!=\!\! 1
\end{equation}

\section{Definition and Formulation of Optimal Topology Transition}


Potential ineffectiveness of the ad hoc topology transition calls for a particular procedure to guarantee transition feasibility, and more importantly, increasingly frequent topology optimization motivates improvement of the performance of topology transition. To this end, in this section, we define the problem of OTT and formulate it mathematically.

\subsection{Definition}

\begin{definition}[Optimal topology transition]\label{def-6-1-1}
  Given values of $\mathcal{P}_p$, an initial topology $\mathcal{E}_0$ and a terminal topology $\mathcal{E}_T$ which both guarantee the feasibility of (\ref{eq-6-1-1:2}) and (\ref{eq-6-1-1:4}), find a topology transition trajectory $S = (\mathcal{E}_0, \mathcal{E}_1,..., \mathcal{E}_{T - 1}, \mathcal{E}_T )$ with $\mathcal{E}_t \subseteq \mathcal{E}_a$, $\forall t \in \{1,2,...,T-1\}$ and satisfying $S \in \mathbb{S}(\mathcal{P}_p)$ such that graphs $\{ \mathcal{G}_t(\mathcal{V}, \mathcal{E}_t) | t = 0, 1,...,T\}$ minimize the transition performance metric $H(S|\mathcal{P}_p)$.
\end{definition}

\begin{figure}[h]
	\centering
	\includegraphics[width=0.88\columnwidth]{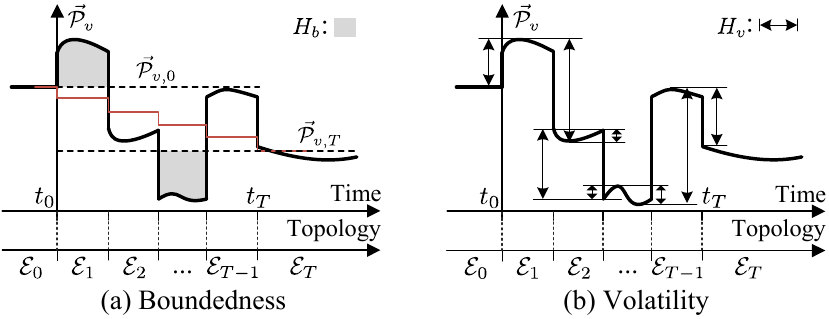}
	\caption{Illustration of boundedness and volatility metrics.}
  \label{fig-6-1-2}
\end{figure}

In Definition \ref{def-6-1-1}, the feasible region $\mathbb{S}$ is shaped by $\mathcal{E}_{t} \!\supset\! \mathcal{E}_u$, Condition \ref{cond-6-1-1} to Condition \ref{cond-6-1-2-x} and switching modes. The transition performance metric $H(S|\mathcal{P}_p)$ should take account of at least two aspects of the transition: transition cost and well-shapedness of system state trajectories in the during-transition stage. The transition cost is the total cost of all switching actions to realize transition. We proposed two metrics to evaluate the well-shapedness. The first is the boundedness metric as shown in Fig. \ref{fig-6-1-2} (a) that indicates how well the system state trajectory in the during-transition stage is bounded by the intersection of all surfaces that pass through the points of system state under topology $\mathcal{E}_0$ and $\mathcal{E}_T$, and are parallel to one axis. More formally, the boundedness metric is defined as
  \begin{equation}\label{eq-6-1-6}
    H_b \!=\!\! \sqrt{ \scaleint{4ex}_{t_0}^{t_T} \!\!  \Delta \vec{\mathcal{P}}_v(t)^T {\bm{w}}_b\D \Delta \vec{\mathcal{P}}_v(t) \mathrm{d} t }
  \end{equation}
  with 
  $
    \Delta \vec{\mathcal{P}}_v(t) = I(\vec{\mathcal{P}}_v(t) \!-\! \vec{\mathcal{P}}_{v,\max}) \!+\! I(\vec{\mathcal{P}}_{v, \min} \!-\! \vec{\mathcal{P}}_v(t))
  $.
  Here $\vec{\mathcal{P}}_{\!v\!,\max} \!\!=\!\! \max( \vec{\mathcal{P}}_{v\!,0}, \vec{\mathcal{P}}_{\!v,T})$ and $\vec{\mathcal{P}}_{v,\min} \!\!=\!\! \min( \vec{\mathcal{P}}_{v,0}, \vec{\mathcal{P}}_{\!v,T})$; $\mathcal{P}_v(t)$ denotes $\mathcal{P}_v$ as the function of time $t$. 
  The second is the volatility metric as shown in Fig. \ref{fig-6-1-2} (b) that evaluates the degree of sudden changes of the system state in the during-transition stage. Mathematically, the volatility metric is 
  \begin{equation}\label{eq-6-1-7}
    H_v \!\!=\!\!  \!
    \sum\nolimits_{k = 2}^{n_K} \! \Vert {\bm{w}}_v\D [ \vec{\mathcal{P}}_v(t_k) \!-\! \vec{\mathcal{P}}_v(t_{k\!-\!1})] \Vert_1  
    \!-\! \Vert \!{\bm{w}}_v\D\! ( \vec{\mathcal{P}}_{v,0} \!-\! \vec{\mathcal{P}}_{v,T}) \Vert_1
  \end{equation}
  where $\{t_k |_{k=1}^{n_K}, t_k \!>\! t_{k-1} \}$ is the set of time during $[t_0^-, t_T^+]$ for which $t_1 = t_0^-$, $t_{n_K} = t_T^+$, and $\vec{\mathcal{P}}_v(t_k)$ for $2 \leq k \leq n_K$ is an extremum of $\vec{\mathcal{P}}_v(t)$. 
  The ideal system state trajectories in the during-transition stage are like the red trajectory in Fig. \ref{fig-6-1-2} (a) which is completely bounded by the intersection just mentioned, and contains only necessary sudden changes of the system state, i.e., $\sum\nolimits_{k = 2}^{n_K} \! \Vert {\bm{w}}_v\D [ \vec{\mathcal{P}}_v(t_k) \!-\! \vec{\mathcal{P}}_v(t_{k\!-\!1})] \Vert_1 = \Vert \!{\bm{w}}_v\D\! ( \vec{\mathcal{P}}_{v,0} \!-\! \vec{\mathcal{P}}_{v,T}) \Vert_1$. The  boundedness and volatility metrics together measure the similarity between a system state trajectory in the during-transition stage and the ideal ones.


\begin{remark}
    For the case where $\mathcal{P}_{vc} \neq \emptyset$, since $\mathcal{P}_{vc}$ is fixed in the during- and post-transition stages, for simplicity, we can remove the elements in $\mathcal{P}_{vc}$ from $\mathcal{P}_{c}$ when formulating the OTT problem. For example, for the DC OTS given by (\ref{eq-6-1-1-add}), $\mathcal{P}_v$ can be reduced to $\{ \bm{p}_{l}, \bm{\theta} \}$.
\end{remark}

\begin{remark}
    The OTT problem is different from the multi-time slot TOPs such as those in \cite{4-1003} and \cite{4-1365}, at least in the performance to be optimized, the coupling constraints between adjacent topologies (Condition \ref{cond-6-1-1} and Condition \ref{cond-6-1-2-x} for the OTT problem), and if the total number of topologies is known beforehand. In fact, the OTT problem is also applicable to any two adjacent topologies given by the multi-time slot TOPs. 
    Moreover, a sequence of topologies from an initial topology to the optimal topology are also generated in some solution approaches to TOPs \cite{4-848, 4-847, 4-838}. However, the main purpose of these topologies is to find the optimal solution of the TOP instead of forming a feasible or legitimate transition trajectory.
\end{remark}

\subsection{Formulation} 
 
\subsubsection{\textbf{Handling of unknown T}}
One difficulty of the OTT problem is that $T$ is optimized implicitly since the total number of transitional topologies is unknown beforehand. 
For the practicality of the OTT formulation, we introduce a given upper bound of $T$, denoted as $T_u$. Then the optimization of $T$ is embedded in the OTT formulation by augmenting transition trajectory $S$ into $(\mathcal{E}_0, \mathcal{E}_1,..., \mathcal{E}_{T\!-\! 1}, \mathcal{E}_{T}, \mathcal{E}_{T\!+\! 1}, ..., \mathcal{E}_{T_u} )$ satisfying
\begin{equation}\label{eq-6-1-6-x1}
  \begin{aligned}
    & \mathcal{E}_{t} \neq \mathcal{E}_{t-1}, \forall t  \llbracket 1, T \rrbracket;  \mathcal{E}_{t} = \mathcal{E}_{t-1}, \forall t \in \llbracket T+1, T_{u} \rrbracket 
  \end{aligned}
\end{equation}
where $\mathcal{E}_{T+ 1}$ to $\mathcal{E}_{T_u}$ are the introduced redundant transitional topologies. 
Although $T$ can be infinite if repeated sequence of line switching actions are allowed, bounding $T$ properly does not change the yielded optimal topology transition trajectory since the optimal value of $T$ is finite. When $T$ reaches a certain threshold, additional line switching actions will neither eliminate violations of operational constraints nor reduce the boundedness metric $H_b$, but will certainly increase the volatility metric $H_v$ and transition cost.

To formulate (\ref{eq-6-1-6-x1}) with $\bm{z}_t$, we first introduce an auxiliary vector $\bm{\delta} \!=\! \col\{ {\delta}_t \!\in\! \mathbb{B} |_{t=1}^{T_u}  \} $ to indicate if two adjacent transitional topologies are identical, which satisfies
\begin{equation}\label{eq-6-1-6-x2} 
    \bm{1}^T | \bm{z}_{t} \!-\! \bm{z}_{t-1}|_{\circ} \!\leq\! M  {\delta}_t,
     \bm{1}^T | \bm{z}_{t} \!-\! \bm{z}_{t-1}|_{\circ}  \!\geq\! {\delta}_t 
     ~~ \forall t \!\in\! \llbracket1, T_u \rrbracket
\end{equation} 
such that ${\delta}_t \!=\! 0$ if $\mathcal{E}_t \!=\! \mathcal{E}_{t-1}$ and ${\delta}_t \!=\! 1$ otherwise. Then condition (\ref{eq-6-1-6-x1}) can be ensured by excluding all impossible cases of sameness between adjacent topologies, i.e., 
$\forall t \!\in\! \llbracket1, T_u -2 \rrbracket $, $({\delta}_t, {\delta}_{t+1}, {\delta}_{t+2} ) \notin \{(0,0,1), (0,1,0), (0,1,1),(1,0,1)\}$. Taking $({\delta}_t, {\delta}_{t+1}, {\delta}_{t+2} ) \!\neq\! (0,0,1)$ for example, this inequality is equivalent to ${\delta}_t + {\delta}_{t+1} + (1 - {\delta}_{t+2}) \!\geq\! 1$ since ${\delta}_t \!+\! {\delta}_{t+1} \!+\! (1 \!-\! {\delta}_{t+2})$ is an integer and has a unique minimum 0 when $({\delta}_t, {\delta}_{t+1}, {\delta}_{t+2} ) = (0,0,1)$. Similarly, condition (\ref{eq-6-1-6-x1}) is equivalently ensured by 
\begin{equation}\label{eq-6-1-6-x3}
  \begin{aligned}
    & {\delta}_t \!+\! {\delta}_{t+1} \!+\! (1 \!-\! {\delta}_{t+2})  \!\geq\! 1 ~~ \forall t \!\in\! \llbracket1, T_u -2 \rrbracket \\
    & {\delta}_t \!+\! (1 \!-\! {\delta}_{t+1}) \!+\! {\delta}_{t+2}  \!\geq\! 1 ~~ \forall t \!\in\! \llbracket1, T_u -2 \rrbracket \\
    & {\delta}_t \!+\! (1 \!-\! {\delta}_{t+1}) \!+\! (1\!-\!\!{\delta}_{t+2})  \!\geq\! 1 ~~ \forall t \!\in\! \llbracket1, T_u -2 \rrbracket \\
    & (1 \!-\! {\delta}_t) \!+\! {\delta}_{t\!+\!1} \!+\! (1\!-\!{\delta}_{t+2})  \!\geq\! 1 ~~ \forall t \!\in\! \llbracket1, T_u -2 \rrbracket
  \end{aligned}
\end{equation}
After the augmentation, the terminal topology $\bm{z}_{T_u} = \bm{z}_{T}$. 

\subsubsection{\textbf{Transition performance metric}} 
A combination of transition cost and the boundedness and volatility metrics, as explained in Section IV-A, is used for the evaluation of transition performance. Analogously to derivation of Condition \ref{cond-6-1-2}, it is assumed that for elements of $\vec{\mathcal{P}}_v$ being concerned in $H_b$ and $H_v$, their during-transition trajectories under each transitional topology can be approximated by their steady-state values under transitional topologies. Also, for derivation of $H_b$ and $H_v$, all lines in the same line switching batch are assumed to be switched synchronously since duration of intermediate systems is generally much shorter than that of transitional systems. 
Hence, $H_b$ and $H_v$ are reformulated as 
\begin{equation}\label{eq-6-1-6-appr}
  H_b \!\!=\!\! \sqrt{ \! \sum\nolimits_{t = 1}^{T_u \!-\!1}\!\! d_t  \Delta \vec{\mathcal{P}}_{v, t}^T {\bm{w}}_b\D \Delta \vec{\mathcal{P}}_{v, t}  } ~\text{with}
\end{equation} 
\begin{equation}
  \Delta \vec{\mathcal{P}}_{v, t} \!=\! I(\vec{\mathcal{P}}_{v, t} \!-\! \vec{\mathcal{P}}_{v,\max}) \!+\! I(\vec{\mathcal{P}}_{v, \min} \!-\! \vec{\mathcal{P}}_{v, t} ) 
\end{equation}
and 
\begin{equation}\label{eq-6-1-7-appr}
  H_v \!=\! \sum\nolimits_{t = 1}^{T_u}  \Vert {\bm{w}}_v\D ( \vec{\mathcal{P}}_{v,t} - \vec{\mathcal{P}}_{v,t-1} )  \Vert_1
  \!-\! \Vert {\bm{w}}_v\D ( \vec{\mathcal{P}}_{v,0} \!-\! \vec{\mathcal{P}}_{v, T } )  \Vert_1  
\end{equation} 

The transition cost is the sum of switching costs caused by the topology transition, formulated as
\begin{equation}\label{eq-6-1-8c}
  H_c = \sum\nolimits_{t=1}^{T_u} \bm{w}_c^T |\bm{z}_t - \bm{z}_{t - 1}|_{\circ} 
\end{equation}
Then the transition performance metric is
\begin{equation}\label{eq-6-1-8}
  H = \alpha_b H_b + \alpha_v H_v + \alpha_c H_c
\end{equation}

\begin{remark}
  In this work, we adopt the weight parameters satisfying $\alpha_{c} \!\gg\! \max\{ \alpha_b, \alpha_v\}$ to impose the highest priority to transition cost. The minimal transition cost to transition from $\mathcal{E}_0$ to $\mathcal{E}_T$ (not necessarily feasible) is $\bm{w}_c^T |\bm{z}_0 \!-\! \bm{z}_T|_{\circ}$, where only necessary switching actions, i.e, that for $\mathcal{E}_T \triangle \mathcal{E}_0$, are executed. The term $\alpha_{c} H_{c}$ acts as a penalty to prevent unnecessary line switching actions which are allowed only when feasibility of topology transition cannot be ensured with the minimal transition cost.
\end{remark}

\subsubsection{\textbf{Feasible region \texorpdfstring{$\mathbb{S}$}{Lg}}}
$\!\!$The constraint $\!\mathcal{E}_{t} \!\!\supset\!\! \mathcal{E}_u\!$ is formulated as
\begin{equation}\label{eq-6-1-add-1}
  \bm{E}_u \bm{z}_t = \bm{1}_{|\mathcal{E}_u|}  ~~ \forall t \in   \llbracket 1, T_u-1 \rrbracket
\end{equation} 

For Condition \ref{cond-6-1-1}, topology $\mathcal{E}_{t-1} \!-\! (\mathcal{E}_{t} \triangle \mathcal{E}_{t-1})$ is equivalent to, in the vector space, ${\bm{z}}_t\D \bm{z}_{t-1}  $. Then according to \cite[Theorem 1]{4-995-ea}, Condition \ref{cond-6-1-1} can be ensured by the following constraints:
\begin{subequations}\label{eq-6-1-9} 
  \begin{align}
    & \bm{E}^T \bm{\vartheta}_t \!-\! \bm{\rho}_t \!+\! M (\bm{1}_{|\mathcal{E}_a|} \!-\! {\bm{z}}_t\D \bm{z}_{t-1} ) \!\geq 0 ~ \forall t \!\in\! \llbracket 1, T_u \rrbracket  \label{eq-6-1-9:0} \\
    & \bm{E}^T \bm{\vartheta}_t \!-\! \bm{\rho}_t \!-\! M (\bm{1}_{|\mathcal{E}_a|} \!-\! {\bm{z}}_t\D \bm{z}_{t-1} ) \!\leq 0 ~ \forall t \!\in\! \llbracket 1, T_u \rrbracket \label{eq-6-1-9:1} \\
    & - M {\bm{z}}_t\D \bm{z}_{t-1} \leq \bm{\rho}_t \leq M {\bm{z}}_t\D \bm{z}_{t-1}  ~ \forall t \!\in\! \llbracket 1, T_u \rrbracket \label{eq-6-1-9:2} \\
    & \bm{E} \bm{\rho}_t = \bm{c} ~~ \forall t \!\in\! \llbracket 1, T_u \rrbracket \label{eq-6-1-9:3}
  \end{align}
\end{subequations}
where the constant uniquely-balanced vector $\bm{c}$ is defined by \cite[Definition 1]{4-995-ea}.

For Condition \ref{cond-6-1-2} and Condition \ref{cond-6-1-2-x}, let $\mathbb{F}( {\mathcal{P}}_p )$ and $\mathbb{F}_r( 
    {\mathcal{P}}_p )$ denote the feasible region of $( \vec{\mathcal{P}}_v, \bm{z})$ determined by all constraints in (\ref{eq-6-1-1:4}) and relaxed (\ref{eq-6-1-1:4}), respectively. For example, with the DC OTS given by (\ref{eq-6-1-1-add}), $\mathbb{F}( {\mathcal{P}}_p )$ is shaped by constraints (\ref{eq-6-1-1-add:6})-(\ref{eq-6-1-1-add:9}), and $\mathbb{F}_r( {\mathcal{P}}_p )$ is shaped by constraints (\ref{eq-6-1-1-add:6})-(\ref{eq-6-1-1-add:9}) but with bound parameters $\overline{\bm{p}}_l$ and $\overline{\bm{\theta}}_l$ replaced by $\overline{\bm{p}}_{lr}$ and $\overline{\bm{\theta}}_{lr}$ respectively. Note that constraint (\ref{eq-6-1-1-add:10}) is dropped here for simplicity since $\bm{p}_g$ is fixed in the during- and post-transition stages. 
    Then we have 
\begin{equation}\label{eq-6-1-10} 
  ( \vec{\mathcal{P}}_{v, t}, \bm{z}_t ) \in \mathbb{F}( \mathcal{P}_p ) ~~ \forall t \!\in\! \llbracket 1, T_u-1 \rrbracket
\end{equation}
\begin{equation}\label{eq-6-1-10-x}
  \!\!( \vec{\mathcal{P}}_{\!vr, t}, \bm{z}_{t\!-\!1} \!+\! \Delta \bm{z}_t \!) \!\!\in\!\! \mathbb{F}_{\!r}( \mathcal{P}_{\!p}) ~~ \forall t \!\!\in\!\! \llbracket 1, T_u \rrbracket, \Delta \bm{z}_t \!\!\in\!\! \mathbb{V}\!(\bm{z}_{t} \!-\! \bm{z}_{t\!-\!1})
\end{equation}

Topology transition under the AS mode requires extra constraints to be imposed on line switching batches. Specifically, $\forall i \in \llbracket 1, n_a \rrbracket$, we first introduce $\bm{\zeta}_i$ being the value of $\bm{z}$ corresponding to topology $\mathcal{E}^i$. Then the condition to ensure the \textit{nonempty} line switching batch $\mathcal{E}_t \triangle \mathcal{E}_{t-1}$ satisfies the requirement of AS mode, is as follows:
\begin{condition}\label{cond-6-1-3}
  $\exists! i \in \llbracket 1, n_a \rrbracket: \bm{\zeta}_i^T |\bm{z}_{t} - \bm{z}_{t-1}|_{\circ} \neq 0 $. 
\end{condition}

Introducing auxiliary variables $\bm{u}_t \!\in\! \mathbb{B}^{n_a}$ for all $t \!\in\! \llbracket 1, T_u \rrbracket$, 
condition \ref{cond-6-1-3} can be reformulated as 
\begin{subequations}\label{eq-6-1-11} 
  \begin{align}
    & M \! ( \bm{u}_{t} \!\!-\! \bm{1} )  \!\!\leq\!\!\! \begin{bmatrix}\!\!
      (\sum_{i=\!1}^{n_a}\! \bm{\zeta}_i^T  - \bm{\zeta}_1^T) |\bm{z}_{t} \!-\! \bm{z}_{t\!-\!1}|_{\circ} \!\\
       \vdots \\
       (\sum_{i=\!1}^{n_a}\! \bm{\zeta}_i^T - \bm{\zeta}_{n_a}^T) |\bm{z}_{t} \!-\! \bm{z}_{t\!-\!1}|_{\circ}\!
    \end{bmatrix} 
    \!\!\!\leq\!\! 
    M( \bm{1} \!\!-\! \bm{u}_{t}  ), \!
    \\
    & \bm{1}^T \bm{u}_t = 1 ~~ \forall t \in \llbracket 1, T_u \rrbracket
  \end{align}
\end{subequations} 
For any empty line switching batch $\mathcal{E}_t\triangle \mathcal{E}_{t\!-\!1}$ that indicates $\bm{z}_t \!=\! \bm{z}_{t-1}$, constraints (\ref{eq-6-1-11}) are also feasible. Thus constraints (\ref{eq-6-1-11}) should be implemented to all $t \!\in\! \llbracket 1, T_u \rrbracket$ for topology transition under the AS mode. 



\subsubsection{\textbf{Linearization}}
Here nonlinear terms $| \bm{z}_{t} \!-\! \bm{z}_{t-1}|_{\circ}$, ${\bm{z}}_t\D \bm{z}_{t-1}$ and $I(\cdot)$ in (\ref{eq-6-1-6-appr}) are linearized. First for ${\bm{z}}_t\D \bm{z}_{t-1}$, by introducing variables $\bm{z}^b_t \in \mathbb{B}^{|\mathcal{E}_a|}$ imposed by the following constraints:
\begin{equation}\label{eq-6-1-lin-1} 
    \bm{z}^b_t \leq \bm{z}_{t},
    \bm{z}^b_t \leq \bm{z}_{t-1},
    \bm{z}^b_t \geq \bm{z}_{t} + \bm{z}_{t-1} - 1 ~~ \forall t \in \llbracket 1, T_u \rrbracket
\end{equation}
we have
\begin{equation}\label{eq-6-1-lin-2}
    {\bm{z}}_t\D \bm{z}_{t-1} = \bm{z}^b_t 
\end{equation}
Then for $| \bm{z}_{t} - \bm{z}_{t-1}|_{\circ}$, by the equivalence between it and its square,  we have
\begin{equation}\label{eq-6-1-lin-3} 
  | \bm{z}_{t} - \bm{z}_{t-1}|_{\circ} = \bm{z}_{t} + \bm{z}_{t-1} - 2 {\bm{z}}_{t}\D \bm{z}_{t-1} = \bm{z}_{t} + \bm{z}_{t-1} - 2 \bm{z}^b_t
\end{equation}
And for $I(\cdot)$ in (\ref{eq-6-1-6-appr}), it appears in the boundedness metric $H_b$ to be minimized. Thus by introducing variables 
$\bm{\rho}_{0, t}^+ \in \mathbb{R}^{n_p}$, $\bm{\rho}_{0, t}^- \in \mathbb{R}^{n_p}$, $\bm{\rho}_{T, t}^+ \in \mathbb{R}^{n_p}$, $\bm{\rho}_{T, t}^- \in \mathbb{R}^{n_p}$ that satisfy
\begin{equation}\label{eq-6-1-lin-4}
  \begin{aligned}
    & \vec{\mathcal{P}}_{v,t} \!-\! \vec{\mathcal{P}}_{v,\max}  \!=\! \bm{\rho}_{0, t}^+ \!-\! \bm{\rho}_{0, t}^- ~~ \forall i \in \llbracket 1, T_u - 1 \rrbracket \\
     & \vec{\mathcal{P}}_{v, \min} \!-\! \vec{\mathcal{P}}_{v,t} \!=\! \bm{\rho}_{T, t}^+ \!-\! \bm{\rho}_{T, t}^- ~~ \forall i \in \llbracket 1, T_u - 1 \rrbracket \\
    & \bm{\rho}_{0, t}^+ \geq 0, \bm{\rho}_{0, t}^- \geq 0, \bm{\rho}_{T, t}^+ \geq 0, \bm{\rho}_{T, t}^- \geq 0  ~~ \forall i \!\in\! \llbracket 1, T_u \!-\! 1 \rrbracket
  \end{aligned}
\end{equation}
we have 
\begin{equation}\label{eq-6-1-add-3} 
  \!\!\! \min H_b  \!\Leftrightarrow\!  \min \! \sqrt{\!  \sum\nolimits_{t = 1}^{T_u \!-\!1}\!\! d_t  ( \bm{\rho}_{0, t}^+ \!+\! \bm{\rho}_{T, t}^+ )^T {\bm{w}}_b\D ( \bm{\rho}_{0, t}^+ \!+\! \bm{\rho}_{T, t}^+ )  }
\end{equation}


\subsubsection{\textbf{Practical issues}} 
Constraints (\ref{eq-6-1-10-x}) complicate the optimization model in two aspects: the high dimension of constraints mainly caused by cardinality of $\mathbb{V}(\!\bm{z}_{t} \!\!-\!\! \bm{z}_{t\!-\!1}\!)$, and the dependency between this dimension and $\bm{z}_t$ to be optimized. Allowing occasional violations of operational constraints in Condition 3, in constraints (\ref{eq-6-1-10-x}), for each $t$, we only consider the intermediate system which is most likely to violate the operational constraints. Violations caused by intermediate systems which are least likely to violate the operational constraints are ignored, yielding Assumption \ref{asp-6-1-1-x} and \ref{asp-6-1-1-xx}.

\begin{assumption}\label{asp-6-1-1-x}
  For the case where $\mathcal{E}_{t} \!-\! \mathcal{E}_{t-1} \!\!=\!\! \emptyset $ or $\mathcal{E}_{t-1} \!-\! \mathcal{E}_{t} \!=\! \emptyset $, if (\ref{eq-6-1-1:4}) with topology $\mathcal{E}_{t-1}$ and that with topology $\mathcal{E}_{t}$ are both feasible, then $\forall \mathcal{E} \!\subseteq\! \mathcal{E}_{t} \!-\! \mathcal{E}_{t-1}$ and $\mathcal{E}' \!\subseteq\! \mathcal{E}_{t-1} \!-\! \mathcal{E}_{t}$, relaxed (\ref{eq-6-1-1:4}) with topology $\mathcal{E}_{t-1} \cup \mathcal{E}$ and that with $\mathcal{E}_{t-1} \!-\! \mathcal{E}'$ are also feasible. 
\end{assumption}
\begin{assumption}\label{asp-6-1-1-xx}
  For the case where $\mathcal{E}_{t} \triangle \mathcal{E}_{t-1} \neq \emptyset $, if (\ref{eq-6-1-1:4}) with topology $\mathcal{E}_{t-1}$ and that with topology $\mathcal{E}_{t}$ are both feasible, and relaxed (\ref{eq-6-1-1:4}) with topology $\mathcal{E}_{t-1} - \mathcal{E}_{t} \triangle \mathcal{E}_{t-1} $ is feasible, then $\forall \mathcal{E} \subseteq \mathcal{E}_{t} - \mathcal{E}_{t-1}$ and $\mathcal{E}' \subseteq \mathcal{E}_{t-1} - \mathcal{E}_{t}$, relaxed (\ref{eq-6-1-1:4}) with topology $\mathcal{E}_{t-1} \cup \mathcal{E} - \mathcal{E}'$ is also feasible. 
\end{assumption}

With Assumption \ref{asp-6-1-1-x} and \ref{asp-6-1-1-xx}, constraints (\ref{eq-6-1-10-x}) are simplified as
\begin{equation}\label{eq-6-1-prac-1}
  ( \vec{\mathcal{P}}_{vr, t}, {\bm{z}}_t\D \bm{z}_{t-1} ) \!\in\! \mathbb{F}_{r}( \mathcal{P}_{p} ) ~~ \forall t \!\in\! \llbracket 1, T_u \rrbracket
\end{equation} 
where ${\bm{z}}_t\D \bm{z}_{t-1}$ corresponds to the topology $\mathcal{E}_{t-1} \!-\! \mathcal{E}_{t} \triangle \mathcal{E}_{t-1}$ in Assumption \ref{asp-6-1-1-xx} when $\mathcal{E}_{t} \triangle \mathcal{E}_{t-1} \!\neq\! \emptyset $. For the case where $\mathcal{E}_{t} \!-\! \mathcal{E}_{t-1} \!=\! \emptyset $ or $\mathcal{E}_{t-1} \!-\! \mathcal{E}_{t} \!=\! \emptyset $, ${\bm{z}}_t\D \bm{z}_{t-1}$ corresponds to topology $\mathcal{E}_{t-1}$ or $\mathcal{E}_{t}$ and (\ref{eq-6-1-prac-1}) are redundant since ${\bm{z}}_t\D \bm{z}_{t-1}$ is also constrained by $\mathbb{F}$ in (\ref{eq-6-1-10}). These redundant constraints can be reserved since most solvers will automatically remove them before solving. 
It is noted that the simplification of constraints (\ref{eq-6-1-10-x}) can cause a limitation of the proposed OTT model since Assumption \ref{asp-6-1-1-x} and \ref{asp-6-1-1-xx} do not always hold. However, this limitation can be practically acceptable if the probability of violations of the operational constraints in Condition 3 caused by the potential invalidity of these assumptions is small enough. This probability will be numerically evaluated later.

In addition, for practical topology transition, there could be a low probability that the minimal total number of switching actions to ensure topology transition satisfies the operational constraints in Condition \ref{cond-6-1-2} and Condition \ref{cond-6-1-2-x} significantly exceed the lower bound $ \bm{1}^T |\bm{z}_0 - \bm{z}_T|_{\circ} $, or even the feasible region of topology transition is a null set. Seeing that occasional violations of the operational constraints in Condition \ref{cond-6-1-2} and Condition \ref{cond-6-1-2-x} can be allowed, the excessive switching actions can be avoided by bounding the total number of switching actions and relaxing the operational constraints. Specifically, by introducing slack variables $\bm{\xi}_t \in \mathbb{R}^{n_o}$, $\forall t \!\in\! \llbracket 1, T_u-1 \rrbracket$ and $\bm{\xi}_{r, t} \in \mathbb{R}^{n_{r, o}}$, $\forall t \!\in\! \llbracket 1, T_u \rrbracket$, (\ref{eq-6-1-10}) and (\ref{eq-6-1-prac-1}) become
\begin{equation}\label{eq-6-1-add-2}
  \begin{aligned}
    & ( \vec{\mathcal{P}}_{v, t}, \bm{z}_t ) \in \mathbb{F}( \bm{\xi}_t |\mathcal{P}_p ) ~~~ \forall t \!\in\! \llbracket 1, T_u-1 \rrbracket  \\
    & ( \vec{\mathcal{P}}_{vr, t}, \bm{z}_t \bm{z}_{t-1} ) \in \mathbb{F}_r( \bm{\xi}_{r, t} |\mathcal{P}_p ) ~~~ \forall t \!\in\! \llbracket 1, T_u \rrbracket  \\
    & \bm{\xi}_t \!\geq\! \bm{0} ~~~ \forall t \!\in\! \llbracket 1, T_u-1 \rrbracket ,
     \bm{\xi}_{r, t} \!\geq\! \bm{0} ~~~ \forall t \!\in\! \llbracket 1, T_u \rrbracket
  \end{aligned}
\end{equation}
where $\mathbb{F}( \bm{\xi}_t |\mathcal{P}_p )$ represents $\mathbb{F}(\mathcal{P}_p )$ with operational constraints slacked by $\bm{\xi}_t$ and $\mathbb{F}_r( \bm{\xi}_{r, t} |\mathcal{P}_p )$ is analogous. 
The total number of switching actions for topology transition is bounded as
\begin{equation}\label{eq-6-1-ns}
  \sum\nolimits_{t=1}^{T_u} \bm{1}^T |\bm{z}_t - \bm{z}_{t - 1}|_{\circ} \leq \bm{1}^T |\bm{z}_0 - \bm{z}_T|_{\circ} + n_e
\end{equation}
The following penalty for $\bm{\xi}_t$ and $\bm{\xi}_{r, t}$ should be minimized together with $H$ with the highest priority:
\begin{equation}
  \alpha_p H_p = \alpha_p \Big( \sum\nolimits_{t=1}^{T_u - 1} \bm{1}^T \bm{\xi}_t + \sum\nolimits_{t=1}^{T_u} \bm{1}^T \bm{\xi}_{r, t} \Big)
\end{equation}
where 
$\alpha_p \!\gg\! \alpha_{c} $. Hence, topology transition can be always realized by limited line switching actions.

Decreasing the total number of line switching batches can reduce the operational complexity. Thus we also minimize the following term together with $H$ with the lowest priority:
\begin{equation}
  \alpha_n H_n = \alpha_n \bm{1}^T \bm{\delta}
\end{equation}
where $\alpha_n \ll \min\{ \alpha_b, \alpha_v \}$.

\subsubsection{\textbf{Final formulation}}
Finally, combining all the above expressions yields the formulation of OTT as follows: 
\begin{equation}\label{eq-6-1-form}
  \!\!\!\! \begin{aligned}
    \text{var.} &  
    \left[
     \begin{aligned}
       & \bm{\rho}_{0, t}^+ \!\in\! \mathbb{R}^{n_p}, \bm{\rho}_{0, t}^- \!\in\! \mathbb{R}^{n_p}, \bm{\rho}_{T, t}^+ \!\in\! \mathbb{R}^{n_p}, \bm{\rho}_{T, t}^- \!\in\! \mathbb{R}^{n_p},  \\  
       & \vec{\mathcal{P}}_{v, t} \!\!\in\! \mathbb{R}^{n_p}\!, \bm{z}_t \!\in\! \mathbb{B}^{|\mathcal{E}_a|}\!, \bm{\xi}_t \!\in\! \mathbb{R}^{n_o }\!, \forall t \!\in\! \llbracket 1, T_u \!-\! 1 \rrbracket; \\
       & \vec{\mathcal{P}}_{vr, t} \!\in\! \mathbb{R}^{n_p }, \bm{z}^b_t \!\in\! \mathbb{B}^{|\mathcal{E}_a|}, \bm{\xi}_{r, t} \!\in\! \mathbb{R}^{n_{r,o}}, \bm{\vartheta}_t \!\in\! \mathbb{R}^{|\mathcal{V}|}, \\
       & \bm{\rho}_t \!\in\! \mathbb{R}^{|\mathcal{E}_a| }, \bm{u}_t \in \mathbb{B}^{n_a}, \forall t \in \llbracket 1, T_u \rrbracket; \bm{\delta} \!\in\! \mathbb{B}^{T_u} 
     \end{aligned}
     \right]  
   \\
    \min &~  
    H'(\bm{\rho}_{0, t}^+, \bm{\rho}_{T, t}^+, \vec{\mathcal{P}}_{v, t}, \bm{z}_t, \bm{z}_t^b, \bm{\delta}, \bm{\xi}_{t}, \bm{\xi}_{r, t} )
    \\ 
    \text{s.t.} &~  
      (\ref{eq-6-1-6-x2})^L, (\ref{eq-6-1-6-x3}), (\ref{eq-6-1-add-1}), (\ref{eq-6-1-9})^L, (\ref{eq-6-1-lin-1}), (\ref{eq-6-1-lin-4}), (\ref{eq-6-1-ns})^L, (\ref{eq-6-1-11}), (\ref{eq-6-1-add-2})^L    
  \end{aligned} \!\!\!
\end{equation}
where $H'(\cdot)$ is the objective function formulated as
\begin{equation}
    \begin{aligned}
        & H'(\cdot) = H + \alpha_p H_p + \alpha_n H_n  \\
        & = \alpha_b \sqrt{\!  \sum\nolimits_{t = 1}^{T_u \!-\!1}\!\! d_t  ( \bm{\rho}_{0, t}^+ \!+\! \bm{\rho}_{T, t}^+ )^T {\bm{w}}_b\D ( \bm{\rho}_{0, t}^+ \!+\! \bm{\rho}_{T, t}^+ )  } \\
        &~~~ + \alpha_v \! \left( \sum\nolimits\limits_{t = 1}^{T_u} \!  \Vert {\bm{w}}_v\D ( \vec{\mathcal{P}}_{\!v,t} \!-\!\! \vec{\mathcal{P}}_{\!v,t-\!1} \!)  \Vert_1 \!\!-\! \Vert {\bm{w}}_v\D ( \vec{\mathcal{P}}_{\!v,0} \!\!-\!\! \vec{\mathcal{P}}_{\!v, T } \!)  \Vert_1 \!\! \right) \\
        &~~~  + \alpha_c \sum\nolimits_{t=1}^{T_u} \bm{w}_c^T (\bm{z}_{t} + \bm{z}_{t-1} - 2 \bm{z}^b_t)  \\
        &~~~  + \alpha_p \Big( \sum\nolimits_{t=1}^{T_u - 1} \bm{1}^T \bm{\xi}_t + \sum\nolimits_{t=1}^{T_u} \bm{1}^T \bm{\xi}_{r, t} \Big) + \alpha_n \bm{1}^T \bm{\delta}
    \end{aligned}
\end{equation}
with $H$ expanded by substituting (\ref{eq-6-1-add-3}), (\ref{eq-6-1-7-appr}), and (\ref{eq-6-1-8c}) with $|\bm{z}_t - \bm{z}_{t - 1}|_{\circ}$ linearized by (\ref{eq-6-1-lin-3}), into (\ref{eq-6-1-8}); $(\cdot)^L$ represents the associated constraints with $\bm{z}_t\D \bm{z}_{t - 1}$ or $|\bm{z}_t - \bm{z}_{t - 1}|_{\circ}$ linearized by (\ref{eq-6-1-lin-2}) or (\ref{eq-6-1-lin-3}), respectively. It is noted that $\Vert {\bm{w}}_v\D ( \vec{\mathcal{P}}_{v,0} \!-\! \vec{\mathcal{P}}_{v, T })  \Vert_1$ in the objective function is constant, and constraints (\ref{eq-6-1-11}) in (\ref{eq-6-1-form}) should be dropped for OTT under the SS mode.

For the TOP formulated as an MICP, the corresponding OTT formulation (\ref{eq-6-1-form}) will be also a tractable MICP. 
For example, for the OTT model implemented in the context of DC OTS given by (\ref{eq-6-1-1-add}), 
with $\bm{p}_{g}$ fixed at its optimum $\bm{p}_{g}^*$, $\mathcal{P}_v$ reduced to $\{ \bm{p}_{l}, \bm{\theta} \}$, and constraint (\ref{eq-6-1-1-add:10}) dropped when forming $\mathbb{F}$ and $\mathbb{F}_r$, (\ref{eq-6-1-add-2}) in (\ref{eq-6-1-form}) is specified as 
\begin{subequations}\label{eq-6-1-dc}
    \begin{align} 
    & 
    -M( \bm{1} - \bm{z}_t)  \leq  {\bm{b}}_l\D \bm{E}^T \bm{\theta}_t  - \bm{p}_{l, t}  \leq M( \bm{1} - \bm{z}_t )   \label{eq-6-1-dc-1}\\[-0.6mm]
    & 
    \bm{E} \bm{p}_{l, t} = \bm{p}_g^* - \bm{p}_d \label{eq-6-1-dc-1-add}\\[-0.6mm]
    & 
    \begin{bmatrix}
      M(\bm{z}_t \!-\!  \bm{1} ) \!-\! \overline{\bm{\theta}}_l \\[-0.6mm]
      - {\overline{\bm{p}}}_l\D  \bm{z}_t
    \end{bmatrix}
    \!\!-\! \bm{\xi}_t
    \!\!\leq\!\!
    \begin{bmatrix}
      \bm{E}^T \! \bm{\theta}_t\\[-0.6mm]
      {\bm{p}}_{l, t}
    \end{bmatrix} 
    \!\!\leq\!\!
    \begin{bmatrix}
        M( \bm{1} \!\!-\! \bm{z}_t ) \!+\! \overline{\bm{\theta}}_l \\[-0.6mm]
        {\overline{\bm{p}}}_l\D  \bm{z}_t
      \end{bmatrix}
      \!\!+\! \bm{\xi}_t
    \label{eq-6-1-dc-3}\\[-0.6mm]
    & 
    -M( \bm{1} - \bm{z}^b_t)  \leq  {\bm{b}}_l\D \bm{E}^T \bm{\theta}_{r, t}  - \bm{p}_{lr, t}  \leq M( \bm{1} - \bm{z}^b_t )   \label{eq-6-1-dc-4} \\[-0.6mm]
    & 
    \bm{E} \bm{p}_{lr, t} = \bm{p}_g^* - \bm{p}_d  \\[-0.6mm]
    & 
    \begin{bmatrix}
        \!M(\bm{z}^b_t \!\!-\!  \bm{1} ) \!\!-\! \overline{\bm{\theta}}_{lr} \!\! \\[-0.6mm]
        - {\overline{\bm{p}}}_{lr}\D  \bm{z}^b_t
    \end{bmatrix}
    \!\!\!-\! \bm{\xi}_{r, t}
    \!\!\leq\!\!\!
    \begin{bmatrix}
        \!\bm{E}^T \! \bm{\theta}_{r, t} \!\! \\[-0.6mm]
        {\bm{p}}_{lr, t}
    \end{bmatrix} 
    \!\!\!\leq\!\!\!
    \begin{bmatrix}
        \!M(\! \bm{1} \!\!-\! \bm{z}^b_t ) \!\!+\! \overline{\bm{\theta}}_{lr} \!\! \\[-0.6mm]
        {\overline{\bm{p}}}_{lr}\D  \bm{z}^b_t
        \end{bmatrix}
        \!\!\!+\! \bm{\xi}_{r, t}
    \label{eq-6-1-dc-6} \\[-0.6mm]
    & M \bm{1}_2 \otimes \bm{z}_t - \bm{\xi}_t \geq \bm{0},  \bm{\xi}_t \geq \bm{0}  \label{eq-6-1-dc-7} \\[-0.6mm]
    & M \bm{1}_2 \otimes \bm{z}^b_t - \bm{\xi}_{r, t} \geq \bm{0}, \bm{\xi}_{r, t} \geq \bm{0} \label{eq-6-1-dc-8} 
    \end{align}
\end{subequations}
where (\ref{eq-6-1-dc-1})-(\ref{eq-6-1-dc-3}) and (\ref{eq-6-1-dc-7}) are for all $t \in \llbracket 1, T_u - 1 \rrbracket$, (\ref{eq-6-1-dc-4})-(\ref{eq-6-1-dc-6}) and (\ref{eq-6-1-dc-8}) are for all $t \in \llbracket 1, T_u \rrbracket$; 
$\bm{\theta}_t$ and $\bm{p}_{l, t}$ constitute $\vec{\mathcal{P}}_{v, t}$; $\bm{\theta}_{r, t}$ and $\bm{p}_{lr, t}$ constitute $\vec{\mathcal{P}}_{vr, t}$; and $n_o \!\!=\!\! n_{r,o} \!\!=\!\! 2|\mathcal{E}_a|$ for slack variables $\bm{\xi}_t$ and $\bm{\xi}_{r, t}$.
Constraints (\ref{eq-6-1-dc-1})-(\ref{eq-6-1-dc-3}) and (\ref{eq-6-1-dc-4})-(\ref{eq-6-1-dc-6}) are both corresponding to (\ref{eq-6-1-1-add:6})-(\ref{eq-6-1-1-add:9}) in the DC OTS model, but respectively for transitional and intermediate systems; and (\ref{eq-6-1-dc-7}) and (\ref{eq-6-1-dc-8}) are constraints for the slack variables, which impose zero values on the slack variables associated with branches which are switched off.

\section{Transition-embedded Topology Optimization and Efficient Solution Algorithm}

\subsection{Transition-Embedded Topology Optimization}
We now consider the strict case where the during-transition system should always satisfy operational constraints and only necessary switching actions are allowed to realize transition. In this case, it is possible that no feasible transition trajectories exist between particular initial and terminal topologies. To address this issue, we further embed the formulation of OTT into TOPs rather than addressing them independently. Hence, transition feasibility is always guaranteed by relaxing the terminal topology.
With the terminal topology unknown, $\vec{\mathcal{P}}_{v, \max}$ and $\vec{\mathcal{P}}_{v, \min}$ also become variables determined by equalities with $\max$ and $\min$ operators. By introducing binary variables $\bm{\eta}_1, \bm{\eta}_2, \bm{\eta}_3, \bm{\eta}_4 \in \mathbb{B}^{n_p}$, these equalities can be linearized as
\begin{subequations}\label{eq-6-1-maxmin}
  \begin{align}
    & -M( \bm{1} - \bm{\eta}_1 ) \leq \vec{\mathcal{P}}_{v, \max} - \vec{\mathcal{P}}_{v, 0}   \leq M( \bm{1} - \bm{\eta}_1 )  \\
    & -M( \bm{1} - \bm{\eta}_2 ) \leq \vec{\mathcal{P}}_{v, \max} - \vec{\mathcal{P}}_{v, T_u} \leq M( \bm{1} - \bm{\eta}_2 ) \\
    & \vec{\mathcal{P}}_{v, T_u} \!\leq\! \vec{\mathcal{P}}_{v, 0} \!+\! M( \bm{1} \!-\! \bm{\eta_1} ),
      \vec{\mathcal{P}}_{v, 0} \!\leq\! \vec{\mathcal{P}}_{v, T_u} \!+\! M( \bm{1} \!-\! \bm{\eta_2} ) \\
    & \bm{\eta}_1 + \bm{\eta}_2 = \bm{1},
      \vec{\mathcal{P}}_{v, \min} + \vec{\mathcal{P}}_{v, \max} = \vec{\mathcal{P}}_{v, 0} + \vec{\mathcal{P}}_{v, T_u} 
  \end{align} 
\end{subequations}
Then we have the formulation of transition-embedded TOPs as follows: 
\begin{subequations}\label{eq-6-1-combine}
  \begin{align}
    \text{var.} &   
    \left[
     \begin{aligned}
       & \text{Variables in}~ (\ref{eq-6-1-form}) ~\text{except for}~ \bm{\xi}_t ~\text{and}~ \bm{\xi}_{r, t},  \\[-0.6mm]
       & \vec{\mathcal{P}}_{\!v, 0} \!\!\in\!\! \mathbb{R}^{n_p}\!, \vec{\mathcal{P}}_{\!v, T_u} \!\!\in\!\! \mathbb{R}^{n_p}\!, \vec{\mathcal{P}}_{\!v, \max} \!\!\in\!\! \mathbb{R}^{n_p}\!, \vec{\mathcal{P}}_{\!v, \min} \!\!\in\!\! \mathbb{R}^{\!n_p}\!, \\[-0.6mm]
       & \bm{z}_{T_u} \!\!\in\! \mathbb{B}^{|\mathcal{E}_a|}\!, \bm{\eta}_1 \!\!\in\! \mathbb{B}^{n_p}\!, \bm{\eta}_2 \!\!\in\! \mathbb{B}^{n_p}\!, \bm{\eta}_3 \!\!\in\! \mathbb{B}^{n_p}\!, \bm{\eta}_4 \!\!\in\! \mathbb{B}^{n_p}
     \end{aligned}
     \! \right]
   \\[-0.6mm]
    \min &~
      \beta f( \vec{\mathcal{P}}_{v, T_u}, \bm{z}_{T_u} | \bm{z}_0, \mathcal{P}_p ) + H + \alpha_n H_n  
    \\[-0.6mm]
    \text{s.t.} &  
        (\ref{eq-6-1-6-x2})^L, (\ref{eq-6-1-6-x3}), (\ref{eq-6-1-add-1}), (\ref{eq-6-1-9})^L, (\ref{eq-6-1-lin-1}), (\ref{eq-6-1-lin-4}), (\ref{eq-6-1-ns})^L, (\ref{eq-6-1-11}), (\ref{eq-6-1-maxmin}) \nonumber \\[-0.6mm]
      & \bm{z}_{T_u} \in \mathbb{E}( \bm{z}_0 ) \label{eq-6-1-combine:c1} \\[-0.6mm]
      & \vec{\mathcal{P}}_{vc, 0} - \vec{\mathcal{P}}_{vc, t} = \bm{0}  ~~ \forall t \in \llbracket 1, T_u \rrbracket \label{eq-6-1-combine:c2} \\
      &  (\vec{\mathcal{P}}_{v, t}, \bm{z}_t ) \in \mathbb{F} ( \mathcal{P}_{p} )  ~~ \forall t \in \llbracket 0, T_u \rrbracket  \label{eq-6-1-combine:c3} \\[-0.6mm]
      & (\vec{\mathcal{P}}_{vr, t}, \bm{z}^b_t) \in \mathbb{F}_r( \mathcal{P}_{p} ) ~~ \forall t \in \llbracket 1, T_u \rrbracket \label{eq-6-1-combine:c4}
    \end{align}
\end{subequations}
where the objective function is the weighted sum of the objective functions of the TOP and OTT problem excluding the penalty component $\alpha_p H_p$ since the during-transition system is required to satisfy operational constraints strictly; 
the positive weight $\beta$ satisfies $\beta \!\gg\! \max\{\alpha_b, \alpha_v, \alpha_c, \alpha_n \}$ to ensure that the objective function of the TOP is optimized with the highest priority; compared to the constraints of the OTT model (\ref{eq-6-1-form}), (\ref{eq-6-1-combine}) includes additional constraints (\ref{eq-6-1-maxmin}) to linearize the equalities with $\vec{\mathcal{P}}_{v, \max}$ and $\vec{\mathcal{P}}_{v, \min}$, (\ref{eq-6-1-combine:c1}) to ensure feasibility of the terminal topology analogously to (\ref{eq-6-1-1:2}), and (\ref{eq-6-1-combine:c2}) to ensure $\mathcal{P}_{vc}$ to be constant in the during- and post-transition stages; constraints (\ref{eq-6-1-combine:c3}) and (\ref{eq-6-1-combine:c4}) are similar to (\ref{eq-6-1-add-2}) in the OTT model (\ref{eq-6-1-form}), but the slack variables are excluded given that the during-transition system is required to satisfy operational constraints strictly, and constraints $(\vec{\mathcal{P}}_{v, t}, \bm{z}_t ) \!\in\! \mathbb{F} ( \mathcal{P}_{p} )$ with $t \!=\! 0$ and $t = T_u$ are additionally included since the associated $\bm{z}_t$ and $\vec{\mathcal{P}}_{v, t}$ are variables. 
Moreover, the term $ - \Vert {\bm{w}}_v\D ( \vec{\mathcal{P}}_{v,0} \!-\! \vec{\mathcal{P}}_{v, T } ) \Vert_1$ in $H_v$ becomes concave but it can be substituted by $ - \bm{w}_v^T (\vec{\mathcal{P}}_{v, \max} - \vec{\mathcal{P}}_{v, \min})$, and the right-hand side of (\ref{eq-6-1-ns}) can be linearized similarly to (\ref{eq-6-1-lin-3}).

Solving (\ref{eq-6-1-combine}) will produce the terminal topology and its transition trajectory simultaneously. When feasible transition trajectories exist for the terminal topology given by (\ref{eq-6-1-1}), solutions of (\ref{eq-6-1-combine}) are the same as that obtained by solving (\ref{eq-6-1-1}) and (\ref{eq-6-1-form}) separately. Otherwise, (\ref{eq-6-1-combine}) can produce a terminal topology which sacrifices its optimality regarding $f$ but guarantees existence of feasible transition trajectories with only necessary switching actions. In the following, we call the terminal topology $\bm{z}_{T}$ obtained by solving (\ref{eq-6-1-1}) the \textit{optimal topology}, and that obtained by solving (\ref{eq-6-1-combine}) the \textit{transition-feasible optimal topology} (TFOT).

\subsection{Efficient Solution Algorithm}
 
The OTT model (\ref{eq-6-1-form}) in form of MICP can be solved directly by some solvers based on the framework of the branch and bound (B\&B) algorithm. Considering that time complexity of the B\&B algorithm is at most $O(2^n)$ with $n$ being the number of binary variables, it is crucial to design problem-specific solution algorithms for the sake of computational efficiency.

Firstly, in (\ref{eq-6-1-form}), binary variables $\bm{z}_t$ and $\bm{z}^b_t$ can considerably impact computational efficiency with increase of the system scale. By (\ref{eq-6-1-lin-2}), $\bm{z}^b_t$ is uniquely determined with given $\bm{z}_t$. Thus we only need to be concerned with $\bm{z}_t$. If only necessary switching actions, i.e, that for $\mathcal{E}_T \triangle \mathcal{E}_0$, are executed, the dimension of $\bm{z}_t$ can be significantly reduced. This can be realized by adding the following constraint to (\ref{eq-6-1-form}): 
\begin{equation}\label{eq-6-1-solution-1}
    \sum\nolimits_{t = 1}^{T_u} (\bm{z}_{t} + \bm{z}_{t-1} - 2 \bm{z}^b_t) = |\bm{z}_0 - \bm{z}_T |_{\circ}^T
\end{equation} 

\begin{algorithm}[t]
    \caption{Solution algorithm for problem (\ref{eq-6-1-form}) }\label{alg-6-1-1}
    \begin{algorithmic}[1]
      
        \State Compute $T_l$ and initialize feasible solution set $\Phi$, $k \!\leftarrow\! 0$
  
        \State $T_u \leftarrow T_l$ \label{alg-6-1-line-1}
  
        \State Solve (\ref{eq-6-1-form}) with (\ref{eq-6-1-solution-1}) and passed feasible solution $S (\Phi)$ \label{alg-6-1-line-a}
  
        \State $\Phi \leftarrow \Phi \cup \{( T^{(k)}, \bm{x}^{(k)}, J^{(k)} )\}  $ \label{alg-6-1-line-b}
      
        \Repeat 
  
        \State $k \leftarrow k + 1 $, $T_u \leftarrow T_u + 2$, the same as line \ref{alg-6-1-line-a} and \ref{alg-6-1-line-b}
  
  
        \Until{$T^{(k)} = T^{(k-1)}$}  \label{alg-6-1-line-2}
  
        \If {$H_p^{(k)} = 0 $}:  \Return $\hat{\bm{z}}^{(k)}$
        \Else:  Repeat lines \ref{alg-6-1-line-1} to \ref{alg-6-1-line-2} but without (\ref{eq-6-1-solution-1}) when solving (\ref{eq-6-1-form})
        \EndIf: \Return  $\hat{\bm{z}}^{(k)}$
  
    \end{algorithmic}
  \end{algorithm}

Secondly, tight estimation of $T_u$ can also remarkably reduce the number of binary variables, which however is far from easy since the optimal value of $T$ can vary widely with changes of $\bm{z}_0$ and $\mathcal{P}_p$. Seeing that solving multiple (\ref{eq-6-1-form}) with smaller $T_u$ can be faster than solving one with larger $T_u$, a more efficient alternative approach is initially setting $T_u$ to the lower bound of $T$ and approaching the optimal value of $T$ progressively. Thirdly, passing a known feasible solution when optimization starts can potentially improve computational efficiency. An accessible feasible solution of (\ref{eq-6-1-form}) is that corresponding to the ad hoc transition $S_{syn}$ or $\!S_{asy}$. Moreover, feasible solutions of (\ref{eq-6-1-form}) with smaller $T_u$ are also feasible to (\ref{eq-6-1-form}) with larger $T_u$.

In light of the above three aspects, an efficient algorithm to solve (\ref{eq-6-1-form}) is designed as given by Algorithm \ref{alg-6-1-1}. Here $T_l$ denotes the lower bound of $T$ and is computed as follows: 
under the SS mode, if $\mathcal{E}_0 \!-\! (\mathcal{E}_0 \triangle \mathcal{E}_T)$ is connected, $T_l \!\!=\!\! 1$ and otherwise  $T_l \!\!=\!\! 2$; and under the AS mode, $T_l$ is the cardinality of $\{ i| i \!\!\in\!\! \llbracket 1, n_a \rrbracket; \mathcal{E}^i \!\cap (\mathcal{E}_0 \triangle \mathcal{E}_T) \!\!\neq\!\! \emptyset \}$. The feasible solution set $\Phi$ is initialized as $\{\!( T, \bm{x}, J )\!\} $ corresponding to the ad hoc transition $S_{syn}$ or $\!S_{asy}$, where $\bm{x}$ collocates all optimization variables and $J$ represents the objective function in (\ref{eq-6-1-form}). The superscript ``$(k)$'' indicates the value of variables after the $k$-th iteration. The function $S(\Phi)$ returns the feasible solution in $\Phi$ with the minimal value of $J$ among those satisfying $T \!\!\leq\!\! T^{(k)}$. Algorithm \ref{alg-6-1-1} is also applicable to solving (\ref{eq-6-1-combine}) with the modification that $\Phi$ is initialized to an empty set.

\section{Case Study}

This section studies the OTT problem numerically. 
For OTT in transmission networks, we consider the IEEE 39-bus and 118-bus systems and the German transmission network, which execute the DC OTS or LPA-based OTS to minimize the dispatch cost. For OTT in distribution networks, we consider the IEEE 33-bus and 123-bus systems and a practical-scale distribution network in Shandong province, China, which execute the DistFlow-based DNR to minimize the power injection from the substations. We use DC-OTS-39, LPA-OTS-39, DC-OTS-118, LPA-OTS-118, DC-OTS-DE, LPA-OTS-DE, DF-DNR-33, DF-DNR-123, DF-DNR-SD to refer to the 8 test cases, respectively. A 400-hour period with a 4-hour execution cycle of OTS or DNR is used. All systems are modified for the testing scenario and detailed data for them is provided in Ref. \cite{4-1002}. All models are implemented in Python using a Linux 64-Bit server with 2 Intel(R) Xeon(R) E5-2640 v4 @ 2.40GHz CPUs and 125GB RAM, where Gurobi 9.1 is used as solvers.

The main points studied here include: 
\textit{(\romannumeral1)} the fragility of ad hoc topology transition to ensure operational feasibility; 
\textit{(\romannumeral2)} the effectiveness of the OTT model to find superior topology transition trajectories compared with the ad hoc ones and the mechanism to achieve this superiority; 
\textit{(\romannumeral3)} the necessity of transition-embedded topology optimization when no operational constraint violations but only necessary switching actions are allowed; 
\textit{(\romannumeral4)} the effectiveness of transition-embedded topology optimization and its impacts on the value of objective function of TOPs; 
and \textit{(\romannumeral5)} the superiority of Algorithm \ref{alg-6-1-1} in computational efficiency compared with directly solving by Gurobi. 
In the following, the OTT problem is first analysed in detail with the results on the first 60 transition scenarios of the DC-OTS-39 case, while complete results on all the cases are then summarized.

\subsection{Results on the DC-OTS-39 Case}

\begin{figure}[ht]
    \centering  
    \subfigure[SS mode.]{\label{fig:a}\includegraphics[width=1\columnwidth]{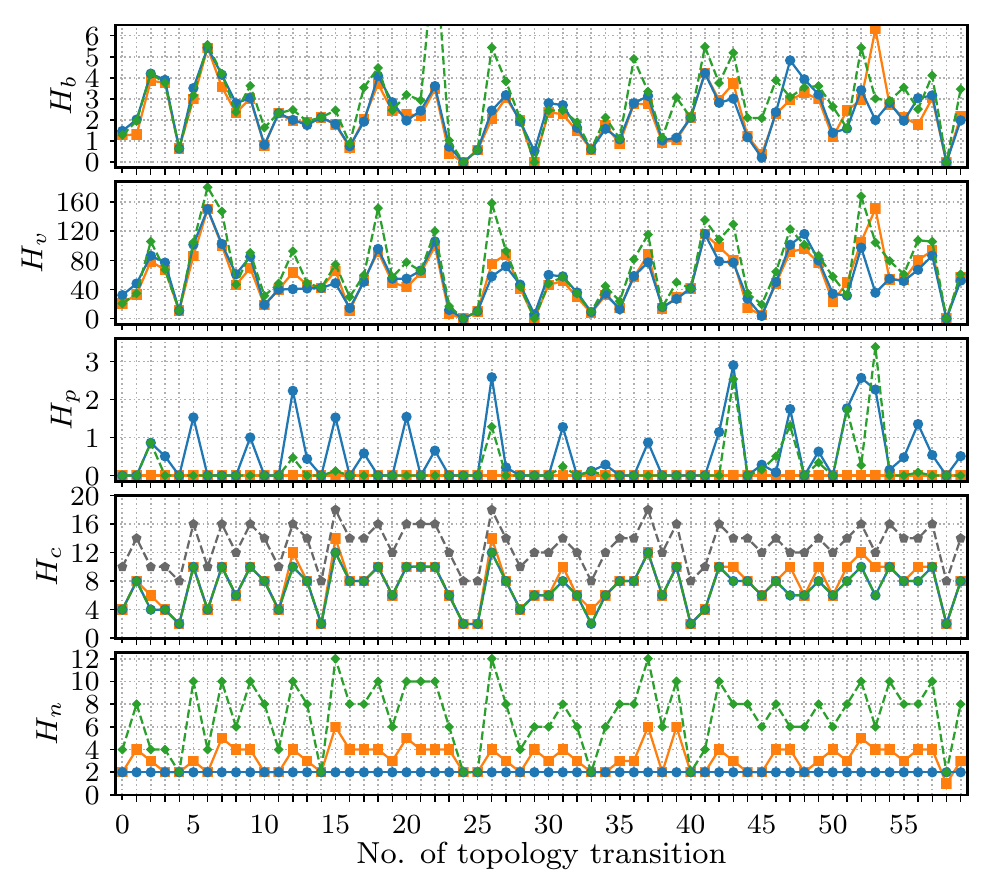}}
    \subfigure[AS mode.]{\label{fig:b}\includegraphics[width=1\columnwidth]{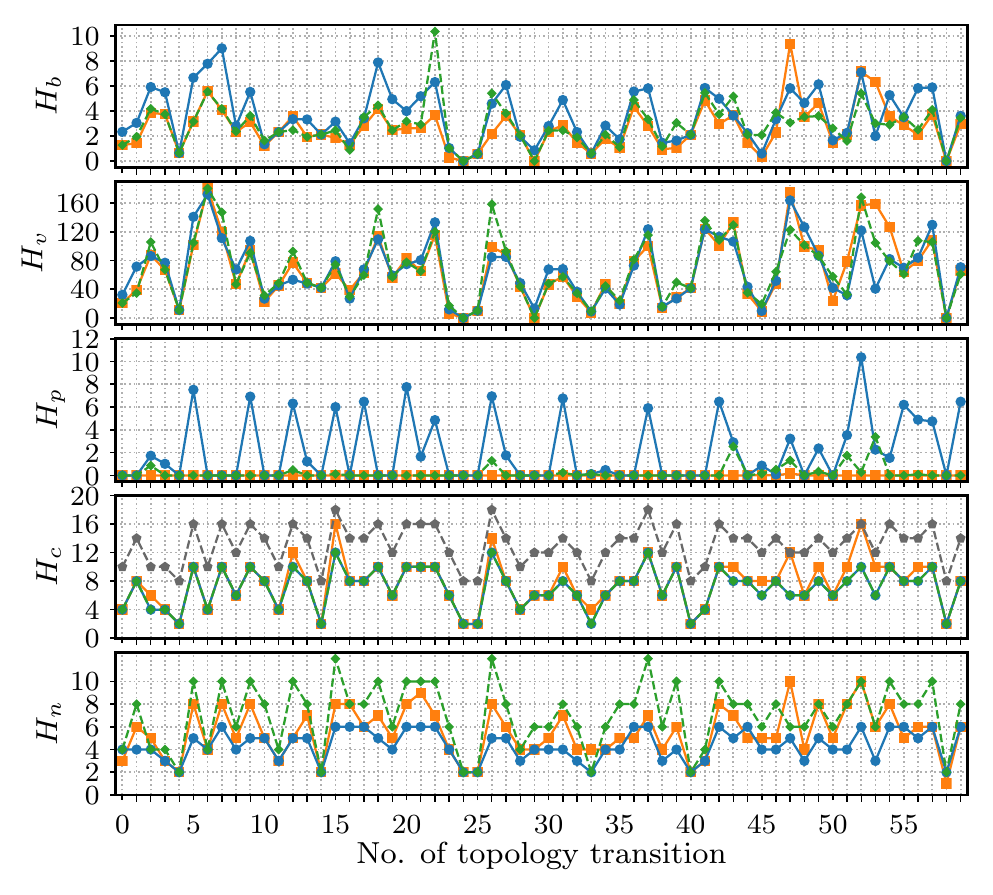}}
    \caption{Comparison of $\{H_*|_{* \in \{b,v,p,c,n\}} \}$ of the ad hoc transition trajectories $S_{syn}$ or $S_{asy}$ (blue points), $S_{one}$ (green points) and optimal transition trajectories given by (\ref{eq-6-1-form}) (orange points), for the DC-OTS-39 case. As shown in the x-axis, all topology transition scenarios are numbered following the execution order of OTS during the 240-hour time period. The gray line for $H_c$ corresponds to the maximal values of $H_c$ determined by (\ref{eq-6-1-ns}).}
    \label{fig-6-1-r12} 
\end{figure}

 Fig. \ref{fig-6-1-r12} compares the transition performance ($H_b$, $H_v$ and $H_c$), violations of operational constraints ($H_p$) and the number of line switching batches ($H_n$) of the ad hoc transition trajectories and the transition trajectory given by (\ref{eq-6-1-form}), for the DC-OTS-39 case. It is noted that $S_{one}$ is also optimized by (\ref{eq-6-1-form}) with $T_u \!=\! \bm{1}^T |\bm{z} \!-\! \bm{z}_T|_{\circ}$ and extra constraint $\bm{1}^T (\bm{z}_{t} + \bm{z}_{t-1} - 2 \bm{z}^b_t) = 1$ for all $t \in \llbracket 1, T_u \rrbracket$. 
According to the results, we can draw the following conclusions:

\textit{(1)} For both the two switching modes, $H_p$ of about half of $S_{syn}$ or $S_{asy}$ is larger than 0, indicating violations of operational constraints in Condition \ref{cond-6-1-2} or Condition \ref{cond-6-1-2-x}. Thus, Presumption \ref{asp-6-1-1} only holds for particular transition scenarios. Especially, under the AS mode, violations of operational constraints of the ad hoc transition trajectories are more severe since $S_{asy}$ contains not only all transitional topologies in $S_{syn}$ but also other transitional topologies to cause possible violations of operational constraints. The ad hoc transition $S_{one}$ performs better than $S_{syn}$ and $S_{asy}$ regarding violations of operational constraints, but it is still found that $H_p > 0$ for a quarter of transition scenarios with $S_{one}$.

\textit{(2)} For the SS mode, the optimal transition trajectories effectively eliminate operational constraint violations for all cases where $H_p \neq 0$ for the ad hoc transition. For the AS mode, there are two optimal transition trajectories, i.e., that for No. 47 and No. 52 topology transition, causing slight operational constraint violations, which can be acceptable if occasional violations are allowed. This difference between these two modes is caused by extra restrictions on the transition process imposed by the AS mode and more severe operational constraint violations under this mode. 

\textit{(3)} For some transition scenarios where operational constraint violations are eliminated, $H_c$ of the ad hoc and optimal transition trajectories are equal. This indicates that the elimination is achieved by optimally allocating the necessary switching actions into more line switching batches than that of the ad hoc transition $S_{syn}$ or $S_{asy}$, or less than that of $S_{one}$. For the others, $H_c$ of the optimal transition trajectories is larger than that of the ad hoc one and the elimination can only be realized by introducing more switching actions except for those necessary. This implies that for more than 20\% transition scenarios (call them \textit{critical scenarios}), if only necessary switching actions are allowed, no feasible transition trajectories exist between the initial and terminal topologies, which further verifies the necessity of transition-embedded topology optimization when no operational constraint violation is allowed.

\textit{(4)} For both of the two switching modes, more than 40\% optimal transition trajectories improve the performance on boundedness $H_b$ and volatility $H_v$ simultaneously compared with that of ad hoc transition $S_{syn}$ or $S_{asy}$. Most of the other transition trajectories causing no such simultaneous improvement achieve elimination of operational constraint violations. In addition, it is also found that improvement of boundedness under the AS mode is more noticeable. More switching batches are required under the AS mode and thus the ad hoc transition without optimization can be more inferior.

\begin{figure}[h]
	\centering
	\includegraphics[width=1\columnwidth]{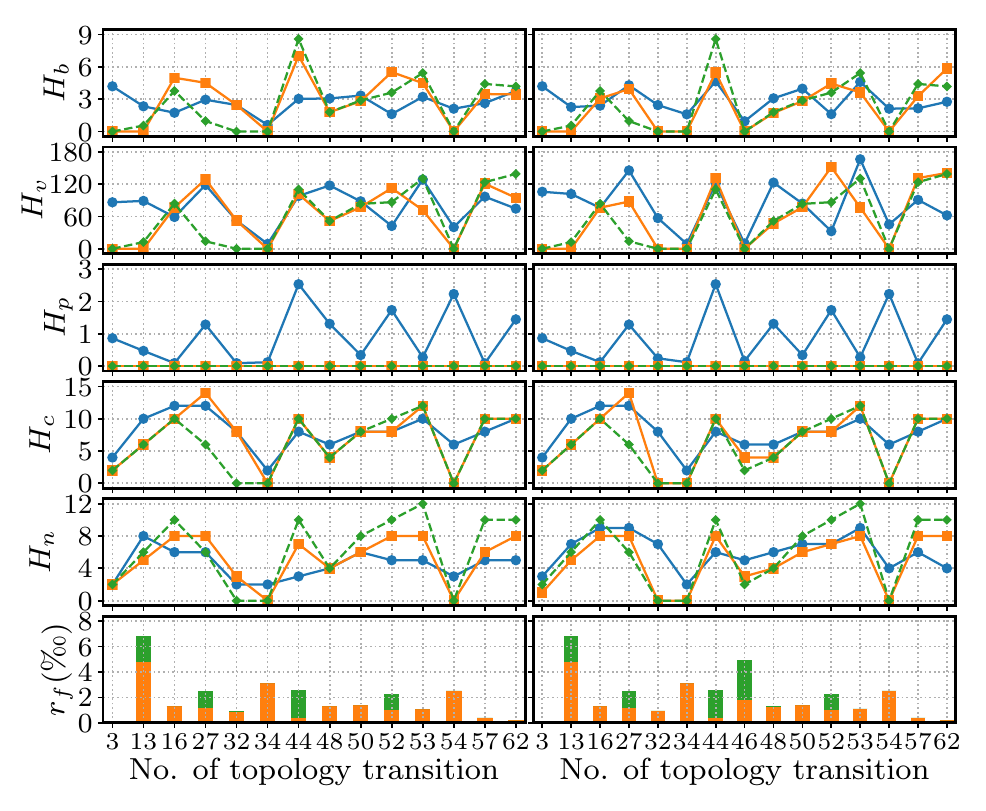}
  \caption{ Comparison of $\{H_*|_{* \in \{b,v,p,c,n\}} \}$ of transition trajectories obtained by solving model 1 (blue points), model 2 (green points) and model 3 (orange points) for critical scenarios of the DC-OTS-39 case. The bar plots compare $r_f$ for the TFOT (orange bars) and ad hoc TFOT (green bars). The left- and right-side figures are results corresponding to SS and AS modes, respectively.}
	\label{fig-6-1-r3} 
\end{figure}

For all critical scenarios, we further solve (\ref{eq-6-1-form}) and (\ref{eq-6-1-combine}) both with $n_e \!=\! 0$, and (\ref{eq-6-1-combine}) with $n_e \!=\! 0$ and extra constraint $\bm{1}^T (\bm{z}_{t} + \bm{z}_{t-1} - 2 \bm{z}^b_t) = {\delta}_t$ for all $t \in \llbracket 1, T_u \rrbracket$ (call them model 1 to 3 respectively). Model 3 is embedding ad hoc transition $S_{one}$ into TOPs and we call its produced optimal topology \textit{ad hoc TFOT} hereinafter. The results are shown in Fig. \ref{fig-6-1-r3}, where $r_f \!=\! (f^*_2 \!-\! f^*_1)\!/\!{f^*_1}$ denotes the dispatch cost change with $f^*_1$ being the optimal dispatch cost of the optimal topology, and $f^*_2$ being that of the TFOT or ad hoc TFOT. By Fig. \ref{fig-6-1-r3}, the following conclusions are drawn:

\textit{(5)} For all critical scenarios, $H_p \!\!>\!\! 0$ for transition trajectories obtained by solving model 1 with the terminal topology obtained by solving (\ref{eq-6-1-1}), while $H_p \!\!=\!\! 0$ for transition trajectories with the terminal topology both obtained by solving model 2. 
In addition, for the TFOT, most values of $r_f$ are less than 2\textperthousand~and all are less than 5\textperthousand. Hence, it is infeasible to transition from the initial topology to the optimal topology with only necessary switching actions for all critical scenarios. In contrast, by slightly increasing the dispatch cost, the TFOT ensures existence of the transition trajectories satisfying operational constraints. 

\textit{(6)} Analogously to the TFOT, the ad hoc TFOT also ensures existence of transition trajectories satisfying operational constraints by increasing the dispatch cost. However, due to ad hoc transition $S_{one}$ limits the switching batch size to 1, for some transition scenarios, e.g., No. 13, the ad hoc TFOT needs to increase more dispatch cost to ensure transition feasibility. 

\textit{(7)} In the No. 3 transition scenario, we can find that $r_f\!\!= H_b\!\!=\!\!H_v\!\!=\!\!H_p\!\!=\!\!0$ for the optimal transition trajectory obtained by solving model 2 or 3. It indicates that for some scenarios, there are two different topologies both with the minimum dispatch cost but only one of them can be transitioned from the initial topology without violating operational constraints and its optimal transition trajectory is with both 0 boundedness and volatility metrics. In this case, only coordinating the TOP and OTT problem can find the superior terminal topology.

\textit{(8)} For some critical scenarios, e.g., No. 34 under either switching mode, the TFOT is the same as the initial topology since $H_c \!\!=\!\! 0$ for its transition trajectory. Hence, reduction of dispatch cost from  topology optimization is possibly unachievable when considering transition feasibility.

In summary, according to the above results on the DC-OTS-39 case, ad hoc topology transition causes violations of operational constraints with considerable probability. In contrast, by optimally allocating the necessary
switching actions into more line switching batches or introducing more switching actions, the optimal transition trajectories yielded by the OTT model eliminate operational constraint violations almost certainly, and is able to improve the performance on boundedness and volatility simultaneously. Moreover, the transition-embedded topology optimization is necessary when no operational constraint violations but only necessary switching actions are allowed. The produced TFOT, with slight increase in dispatch cost, ensures the existence of feasible transition trajectories.

Additionally, the ad hoc topology transition trajectories are possible outputs by simple ordering methods which naively serialize and order switching actions. Thus, the above comparison also demonstrates the necessity of the proposed topology transition method when considering such simple ordering methods. The superiority of the optimal transition trajectories also implies the advantages of the proposed topology transition method over the simple ordering methods.

\begin{table*}[b]
    \caption{Result statistics for all the cases}
    \centering

    \setlength{\tabcolsep}{0pt} 

    \setlength{\aboverulesep}{0pt}
    \setlength{\belowrulesep}{0pt}
    \setlength{\extrarowheight}{.1ex}
    \small
    {
    \begin{tabularx}{1.0\textwidth}{p{2cm}p{0.75cm}<{\centering}p{0.75cm}<{\centering}p{0.75cm}<{\centering}p{0.75cm}<{\centering}p{0.75cm}<{\centering}p{0.35cm}<{\centering}p{1.03cm}<{\centering}p{0.75cm}<{\centering}p{0.75cm}<{\centering}p{0.72cm}<{\centering}p{0.72cm}<{\centering}p{0.7cm}<{\centering}p{0.7cm}<{\centering}p{0.7cm}<{\centering}p{0.75cm}<{\centering}p{0.9cm}<{\centering}p{0.9cm}<{\centering}p{0.72cm}<{\centering}p{0.72cm}<{\centering}p{1.05cm}<{\centering}p{0.79cm}<{\centering}}
      \toprule 
      ~Cases &
      $r_{1s}$ & $r_{1a}$ & $r_{1o}$ & $r_{2s}$ & $r_{2a}$ & $e_{1s}$ & $e_{1a}$ & 
      $r_{3s}$ & $r_{3a}$ & $r_{4s}$ & $r_{4a}$ & $r_{5s}$ & $r_{5a}$ & 
      $r_{6s}$ & $r_{6a}$ & $e_{2s}$ & $e_{2a}$  & $r_{7s}$ & $r_{7a}$ & $\rho_s$ & $\rho_a$ \\
      \midrule
      ~DC-OTS-39 & 
      0.51 &  0.55 &  0.27 &  1.00 &  0.95 &  0 &  0.053 &  
      0.22 &  0.24 &  0.42 &  0.45 &  1.00 &  1.00 &  
      0.82 &  0.83 &  4.8\textperthousand &  4.8\textperthousand &  0.36 &  0.38 & 
      0 & 0.62\textperthousand  \\
      ~DC-OTS-118 &
      0.57 &  0.59 &  0.31 &  1.00 &  0.97 &  0 &  0.029 &  
      0.31 &  0.34 &  0.37 &  0.46 &  1.00 &  1.00 &  
      0.90 &  0.88 &  2.5\textperthousand &  6.1\textperthousand &  0.26 &  0.29 &
      1.13\textperthousand   & 0.45\textperthousand     \\
      ~DC-OTS-DE &
      0.62 &  0.65 &  0.25 &  1.00 &  0.98 &  0 &  0.086 &  
      0.20 &  0.25 &  0.58 &  0.63 &  1.00 &  1.00 &  
      1.00 &  1.00 &  0.7\textperthousand &  0.7\textperthousand &  0.20 &  0.24 &
      0.41\textperthousand  & 0       \\ \midrule
      ~LPA-OTS-39 & 
      0.55 &  0.57 &  0.33 &  1.00 &  0.93 &  0 &  0.021 &  
      0.27 &  0.31 &  0.53 &  0.55 &  1.00 &  1.00 &  
      0.78 &  0.77 &  5.2\textperthousand &  5.7\textperthousand &  0.30 &  0.32 &
      0.89\textperthousand  &  0           \\
      ~LPA-OTS-118 &
      0.56 &  0.62 &  0.31 &  1.00 &  0.98 &  0 &  0.008 &  
      0.35 &  0.40 &  0.41 &  0.47 &  1.00 &  1.00 &  
      0.86 &  0.88 &  3.2\textperthousand &  3.2\textperthousand &  0.34 &  0.33 &
      0.74\textperthousand & 0              \\
      ~LPA-OTS-DE &
      0.67 &  0.73 &  0.28 &  1.00 &  0.98 &  0 &  0.014 &  
      0.24 &  0.25 &  0.52 &  0.60 &  1.00 &  1.00 &  
      0.96 &  0.96 &  2.9\textperthousand &  2.9\textperthousand &  0.17 &  0.16 &
      0.54\textperthousand & 0.40\textperthousand             \\ \midrule
      ~DF-DNR-33 & 
      0.33 &  0.36 &  0.15 &  1.00 &  0.97 &  0 &  0.068 &  
      0.14 &  0.16 &  0.31 &  0.34 &  1.00 &  1.00 &  
      0.86 &  0.81 &  4.4\textperthousand &  5.8\textperthousand &  0.28 &  0.25 & 
       0       &   0                  \\
      ~DF-DNR-123 &
      0.38 &  0.41 &  0.19 &  1.00 &  1.00 &  0 &  0 &  
      0.17 &  0.22 &  0.26 &  0.30 &  1.00 &  1.00 &  
      1.00 &  1.00 &  0.8\textperthousand &  0.9\textperthousand &  0.18 &  0.23 & 
      0.31\textperthousand       &      0           \\
      ~DF-DNR-SD &
      0.44 &  0.49 &  0.24 &  1.00 &  0.96 &  0 &  0.005 &  
      0.20 &  0.23 &  0.37 &  0.42 &  1.00 &  1.00 &  
      1.00 &  1.00 &  1.3\textperthousand &  1.3\textperthousand &  0.30 &  0.30 & 
      0.48\textperthousand  &  0                        \\
      \bottomrule
      \multicolumn{22}{p{0.98\textwidth}}{
        \footnotesize{
        \textit{Note:} All values are rounded to 2 decimal places (DPs), except for values of $e_{1a}$ being rounded to 3 DPs and that of $e_{2s}$, $e_{2a}$, $\rho_s$ and $\rho_a$ to 1 DP.
      }}
      \end{tabularx}
    }
	\label{table-6-1-1}
\end{table*}

\subsection{Results Statistics for All the Cases}

We similarly investigate the other cases by computing the following metrics: 
\begin{itemize}[leftmargin=10pt, labelsep=3pt]
    \item $r_{1s}$ ($r_{1a}$, $r_{1o}$): ratio of $S_{syn}$ ($S_{asy}$, $S_{one}$) with $H_p > 0$;
    \item $r_{2s}$ ($r_{2a}$): ratio of the optimal transition trajectories which eliminate operational constraint violations under the SS (AS) mode to $S_{syn}$ ($S_{asy}$) with $H_p \neq 0$;
    \item $e_{1s}$ ($e_{1a}$): the maximal value of $H_{p2}/H_{p1}$ under the SS (AS) mode, with $H_{p1} \neq 0 $ being the value of $H_p$ of $S_{syn}$ ($S_{asy}$) and $H_{p2}$ the value of $H_p$ of the corresponding optimal transition trajectory;
    \item $r_{3s}$ ($r_{3a}$): ratio of critical scenarios under the SS (AS) mode;
    \item $r_{4s}$ ($r_{4a}$): ratio of the optimal transition trajectories under the SS (AS) mode which improve the performance on $H_b$ and $H_v$ simultaneously compared with that of $S_{syn}$ ($S_{asy}$);
    \item $r_{5s}$ ($r_{5a}$): ratio of critical scenarios under the SS (AS) mode where $H_p >0$ for transition trajectories obtained by solving model 1 with the terminal topology obtained by solving (\ref{eq-6-1-1}), and $H_p=0$ for transition trajectories with the terminal topology both obtained by solving model 2;
    \item $r_{6s}$ ($r_{6a}$): ratio of TFOT under the SS (AS) mode whose $r_f \!<$2\textperthousand;
    \item $e_{2s}$ ($e_{2a}$): the maximal value of $r_f$ of TFOT under the SS (AS) mode;
    \item $r_{7s}$ ($r_{7a}$): ratio of ad hoc TFOT whose value of $r_f$ is larger than that of the corresponding TFOT. 
\end{itemize}

Table \ref{table-6-1-1} lists the results about the above metrics for all the cases, from which the following conclusions regarding the generalization of what we concluded from the DC-OTS-39 case can be drawn:

\textit{(1)} All qualitative conclusions drawn from the DC-OTS-39 case are still valid with changes in the size of networks, model of power flow or type of systems.

\textit{(2)} For larger systems, LPA-based OTS compared with DC OTS, or OTS compared with DNR, values of $r_{1s}$ and $r_{1a}$ are larger overall, indicating a higher probability of violations of operational constraints in Condition \ref{cond-6-1-2} and Condition \ref{cond-6-1-2-x}. In these cases, therefore, OTT is more useful in general.

\textit{(3)} For practical-scale systems, i.e., the German transmission network and the distribution network in Shangdong province, values of $r_{6s}$ and $r_{6a}$ are larger while values of $e_{2s}$ and $e_{2a}$ are notably smaller, compared with those of the IEEE 39-bus system. Therefore, sacrifice of the objective function caused by transition-embedded topology optimization is less significant for these practical systems.

\begin{table}[htbp!]
	\centering
    \caption{Computation time of Algorithm \ref{alg-6-1-1} and directly using Gurobi.} 
    \vspace{-5.4pt}
    \setlength{\tabcolsep}{0pt} 

    \setlength{\aboverulesep}{0pt}
    \setlength{\belowrulesep}{0pt}
    \setlength{\extrarowheight}{-.1ex}
    \begin{tabular*}{\hsize}{p{0.5cm}p{0.8cm}<{\centering}p{0.9cm}<{\centering}p{1.1cm}<{\centering}p{0.75cm}<{\centering}p{0.96cm}<{\centering}p{1.06cm}<{\centering}p{0.75cm}<{\centering}p{0.93cm}<{\centering}p{1.15cm}<{\centering}}\toprule
    & \multicolumn{3}{c}{DC-OTS-} 
    & \multicolumn{3}{c}{LPA-OTS-} 
    & \multicolumn{3}{c}{DF-DNR-}  
    \\
    \cmidrule(l){2-4}\cmidrule(l){5-7}\cmidrule(lr){8-10} 
    & 39 & 118  & DE 
    & 39   & 118  & DE 
    & 33   & 123    & SD   
    \\ \midrule
    $~\!\tau_{1g}$  & \small{18.3}  & \small{152.5  }  & \small{2306.9  } & \small{61.8  } & \small{576.3 }  & \small{9698.1}    &\small{ 13.1 }  &\small{ 103.9  }  & \small{2637.6 } \\
    $~\!\tau_{1a}$  & \small{11.5}  & \small{63.5 ~ }  & \small{559.4 ~ } & \small{42.0  } & \small{293.9 }  & \small{2234.8}    &\small{ 7.4 ~}  &\small{ 48.9 ~ }  & \small{738.5 ~}  \\
    $~\!\tau_{2g}$  & \small{29.6}  & \small{ 199.8 }  & \small{2030.1  } & \small{ 93.9 } & \small{ 726.1}  & \small{7952.5}    &\small{  16.2}  &\small{  167.3 }  & \small{ 3560.7}   \\
    $~\!\tau_{2a}$  & \small{18.1}  & \small{ 111.9 }  & \small{592.0 ~ } & \small{ 63.8 } & \small{ 370.3}  & \small{2569.3}    &\small{  12.2}  &\small{  75.3 ~}  & \small{ 1179.5} \\
    \bottomrule
    \multicolumn{10}{@{}p{1\columnwidth}@{}}{
        \footnotesize{\textit{Note}:
        All values are in seconds.
      }}
    \end{tabular*} 
    \label{tab-6-1-2}  
\end{table}

Additionally, we also compute the following metrics which evaluate the limitation of the OTT model caused by the potential invalidity of Assumption \ref{asp-6-1-1-x} and \ref{asp-6-1-1-xx}: 
\begin{itemize}[leftmargin=10pt, labelsep=3pt]
    \item $\rho_{s} (\rho_{a})$: the probability of violations of the operational constraints in Condition 3 for the optimal transition trajectories under the SS (AS) mode. Taking $p_{s}$ as an example, it is computed as 
    \begin{equation}
        \rho_{s} = \frac{ \sum_{i = 0}^{99} \sum_{t = 1}^{T_u} n_{t, i}  }{  \sum_{i = 0}^{99} \sum_{t = 1}^{T_u} |\mathbb{V}(\bm{z}_{t, i}^* - \bm{z}_{t-1, i}^*)| }
    \end{equation}
    where $\bm{z}_{t, i}^*$ denotes the value of $\bm{z}_{t}$ of the optimal transition trajectory under the SS mode for the No. $i$ transition scenario; and $n_{t, i}$ denotes the number of $\Delta \bm{z}_{t, i}$ which satisfies $\Delta \bm{z}_{t, i} \!\in\! \mathbb{V}(\bm{z}_{t, i}^* - \bm{z}_{t-1, i}^*)$ and relaxed (\ref{eq-6-1-1:4}) with $\bm{z} \!=\! \bm{z}_{t-1, i}^* + \Delta \bm{z}_{t, i}$ is violated.
\end{itemize}

The last two columns of Table \ref{table-6-1-1} show the results of $\rho_{s}$ and $\rho_{a}$ for each case. It can be seen that all values of $\rho_{s}$ and $\rho_{a}$ are very small, almost all of them are smaller than 1\textperthousand, and half of them equal to 0. This indicates that the limitation of the proposed OTT model caused by the potential invalidity of Assumption \ref{asp-6-1-1-x} and \ref{asp-6-1-1-xx} is acceptable in practical applications, given that occasional violations of the operational constraints in Condition 3 are practically allowed.

Lastly, the superiority of Algorithm \ref{alg-6-1-1} is assessed by comparing the average computation time of solving associated models by Algorithm \ref{alg-6-1-1} and directly by Gurobi, as shown in Table \ref{tab-6-1-2}. Here 
$\tau_{1g}$ and $\tau_{1a}$ ($\tau_{2g}$ and $\tau_{2a}$) are the average computation time of solving (\ref{eq-6-1-form}) ((\ref{eq-6-1-combine}) with $n_e \!=\! 0$) directly by Gurobi and by Algorithm \ref{alg-6-1-1}, respectively; all average values are over the 200 transition scenarios including the 100 scenarios under the SS mode and another 100 under the AS mode, except for $\tau_{1g}$ and $\tau_{2g}$ for DC-OTS-DE, LPA-OTS-DE and DF-DNR-SD which are over 50 transition scenarios to make the total testing time acceptable. 
By Table \ref{tab-6-1-2}, it can be found that for all the test cases, the value of $\tau_{1a}$ is much smaller than that of $\tau_{1g}$, and the value of $\tau_{2a}$ is also much smaller than that of $\tau_{2g}$. Thus, the proposed solution algorithm is much more efficient than directly using mixed-integer programming solvers for solving both the OTT and transition-embedded TOP models.

\section{Conclusion}

This paper for the first time proposes the concept of OTT, develops its mathematical formulation and designs a problem-specific solution algorithm. Under certain assumptions, the OTT problem is formulated as a mixed-integer program. For given initial and terminal topologies, the OTT model can find the topology transition trajectory that optimizes transition cost and boundedness and volatility of system states during transition while satisfying operational constraints. The transition-embedded topology optimization produces the optimal topology and its transition trajectory simultaneously. The case studies demonstrate that the proposed OTT model effectively finds satisfactory transition trajectories that overcome the fragility of ad hoc topology transition, and the transition-embedded topology optimization model always ensures existence of feasible transition trajectories.

Future work includes OTT considering system stability, distributed implementation of OTT under the AS mode, OTT using the AC power flow model, fast identification of transition feasibility, and coordinated transition of topology and coordinated properties in $\mathcal{P}_{vc}$ \cite{4-1000}.

\ifCLASSOPTIONcaptionsoff
  \newpage
\fi

\bibliographystyle{IEEEtran}
\bibliography{main}

\end{document}

%% file: Preamble/preamble.tex
\usepackage{booktabs} 
\usepackage{multirow}
\usepackage{mathtools}

\usepackage{amssymb}

\usepackage{enumitem}

\usepackage{smartdiagram}
\usesmartdiagramlibrary{additions}

\graphicspath{{Figs/}}

\usepackage{pdfpages}
\usepackage{pdflscape}
\usepackage{amsmath}
\usepackage{amsthm}

\DeclareMathOperator*{\argmin}{arg\,min}

\DeclareMathOperator*{\col}{col}

\usepackage{amsfonts}

\usepackage{color}

\usepackage{setspace}
\usepackage{enumitem}
\usepackage{xcolor}

\usepackage{bm}
\usepackage{bbm}

\usepackage{tikz}
\usepackage{pgf}
\usetikzlibrary{arrows,automata}
\usetikzlibrary{shapes}
\tikzstyle{data44}=[rectangle split,rectangle split parts=2,draw,text centered]

\usepackage{subfigure}

\DeclareMathAlphabet{\mathcal}{OMS}{cmsy}{m}{n}

\usetikzlibrary{matrix}

\tikzset{
  BarreStyle/.style =   {opacity=.3,line width=14 mm,color=#1},
  node style ge/.style={},
  node style sp/.style={},
  yl/.style={},
  arrow style mul/.style={},
}

\newtheorem{assumption}{Assumption}

\newtheorem{remark}{Remark}
 
\newtheorem{definition}{Definition}
\newtheorem{condition}{Condition} 
\newtheorem{presumption}{Presumption}

\usepackage{threeparttable,booktabs}

\usepackage[colorlinks,
            linkcolor=blue,
            anchorcolor=blue,
            citecolor=blue
            ]{hyperref}

\usepackage{cite}

\usepackage{mdwlist}

\usepackage{algorithm}
\usepackage{algorithmicx}
\usepackage{algpseudocode}

\errorcontextlines\maxdimen

\makeatletter
    \newcommand*{\algrule}[1][\algorithmicindent]{\makebox[#1][l]{\hspace*{.5em}\thealgruleextra\vrule height \thealgruleheight depth \thealgruledepth}}%
\newcommand*{\thealgruleextra}{}
\newcommand*{\thealgruleheight}{.75\baselineskip}
\newcommand*{\thealgruledepth}{.25\baselineskip}

\newcount\ALG@printindent@tempcnta
\def\ALG@printindent{%
    \ifnum \theALG@nested>0
        \ifx\ALG@text\ALG@x@notext
        \else
            \unskip
            \addvspace{-1pt}
            \ALG@printindent@tempcnta=1
            \loop
                \algrule[\csname ALG@ind@\the\ALG@printindent@tempcnta\endcsname]%
                \advance \ALG@printindent@tempcnta 1
            \ifnum \ALG@printindent@tempcnta<\numexpr\theALG@nested+1\relax
            \repeat
        \fi
    \fi
    }%
\usepackage{etoolbox}
\patchcmd{\ALG@doentity}{\noindent\hskip\ALG@tlm}{\ALG@printindent}{}{\errmessage{failed to patch}}
\makeatother

\newbox\statebox
\newcommand{\myState}[1]{%
    \setbox\statebox=\vbox{#1}%
    \edef\thealgruleheight{\dimexpr \the\ht\statebox+1pt\relax}%
    \edef\thealgruledepth{\dimexpr \the\dp\statebox+1pt\relax}%
    \ifdim\thealgruleheight<.75\baselineskip
        \def\thealgruleheight{\dimexpr .75\baselineskip+1pt\relax}%
    \fi
    \ifdim\thealgruledepth<.25\baselineskip
        \def\thealgruledepth{\dimexpr .25\baselineskip+1pt\relax}%
    \fi
    \State #1%
    \def\thealgruleheight{\dimexpr .75\baselineskip+1pt\relax}%
    \def\thealgruledepth{\dimexpr .25\baselineskip+1pt\relax}%
}

\usepackage{textcomp}

\usepackage{stmaryrd}

\usepackage{comment}

\usepackage{graphicx,subfigure}

\usepackage{stfloats}
\usepackage{tabularx}

\newcommand{\D}{^{\diamond}}

\usepackage{scalerel}[2016-12-29]
\def\scaleint#1{\vcenter{\hbox{\scaleto[3ex]{\displaystyle\int}{#1}}}}

%% file: main.bbl
\begin{thebibliography}{10}
\providecommand{\url}[1]{#1}
\csname url@samestyle\endcsname
\providecommand{\newblock}{\relax}
\providecommand{\bibinfo}[2]{#2}
\providecommand{\BIBentrySTDinterwordspacing}{\spaceskip=0pt\relax}
\providecommand{\BIBentryALTinterwordstretchfactor}{4}
\providecommand{\BIBentryALTinterwordspacing}{\spaceskip=\fontdimen2\font plus
\BIBentryALTinterwordstretchfactor\fontdimen3\font minus
  \fontdimen4\font\relax}
\providecommand{\BIBforeignlanguage}[2]{{%
\expandafter\ifx\csname l@#1\endcsname\relax
\typeout{** WARNING: IEEEtran.bst: No hyphenation pattern has been}%
\typeout{** loaded for the language `#1'. Using the pattern for}%
\typeout{** the default language instead.}%
\else
\language=\csname l@#1\endcsname
\fi
#2}}
\providecommand{\BIBdecl}{\relax}
\BIBdecl

\bibitem{4-62}
E.~B. Fisher, R.~P. Oneill, and M.~C. Ferris, ``Optimal transmission
  switching,'' \emph{IEEE Trans. Power Syst.}, vol.~23, no.~3, pp. 1346--1355,
  Aug. 2008.

\bibitem{4-361}
C.~{Li}, H.-D. {Chiang}, and Z.~{Du}, ``Online line switching method for
  enhancing the small-signal stability margin of power systems,'' \emph{IEEE
  Trans. Smart Grid}, vol.~9, no.~5, pp. 4426--4435, Sept. 2018.

\bibitem{4-811}
M.~Baran and F.~Wu, ``Network reconfiguration in distribution systems for loss
  reduction and load balancing,'' \emph{IEEE Trans. Power Deliv.}, vol.~4,
  no.~2, pp. 1401--1407, Apr. 1989.

\bibitem{4-913}
F.~Shariatzadeh, C.~B. Vellaithurai, S.~S. Biswas, R.~Zamora, and A.~K.
  Srivastava, ``Real-time implementation of intelligent reconfiguration
  algorithm for microgrid,'' \emph{IEEE Trans. Sustain. Energy}, vol.~5, no.~2,
  pp. 598--607, Apr. 2014.

\bibitem{4-859}
T.~Han and D.~J. Hill, ``H2-norm transmission switching to improve synchronism
  of low-inertia power grids,'' \emph{{IFAC}-{PapersOnLine}}, vol.~53, no.~2,
  pp. 13\,299--13\,304, Jul. 2020.

\bibitem{4-61}
B.~Kocuk, S.~S. Dey, and X.~A. Sun, ``New formulation and strong misocp
  relaxations for ac optimal transmission switching problem,'' \emph{IEEE
  Trans. Power Syst.}, vol.~32, no.~6, pp. 4161--4170, Nov. 2017.

\bibitem{4-807}
B.~Enacheanu, B.~Raison, R.~Caire, O.~Devaux, W.~Bienia, and N.~HadjSaid,
  ``Radial network reconfiguration using genetic algorithm based on the matroid
  theory,'' \emph{IEEE Trans. Power Syst.}, vol.~23, no.~1, pp. 186--195, Feb.
  2008.

\bibitem{4-1285}
N.~Martins, E.~J. de~Oliveira, W.~C. Moreira, J.~L.~R. Pereira, and R.~M.
  Fontoura, ``Redispatch to reduce rotor shaft impacts upon transmission loop
  closure,'' \emph{IEEE Trans. Power Syst.}, vol.~23, no.~2, pp. 592--600, May
  2008.

\bibitem{4-1350}
R.~B. Duffey and T.~Ha, ``The probability and timing of power system
  restoration,'' \emph{IEEE Trans. Power Syst.}, vol.~28, no.~1, pp. 3--9, Feb.
  2013.

\bibitem{4-1351}
Y.~Liu, R.~Fan, and V.~TERZIJA, ``Power system restoration: a literature review
  from~2006 to 2016,'' \emph{J. Mod. Power Syst. Clean Energy}, vol.~4, no.~3,
  pp. 332--341, Jul. 2016.

\bibitem{4-1352}
D.~Fan, Y.~Ren, Q.~Feng, Y.~Liu, Z.~Wang, and J.~Lin, ``Restoration of smart
  grids: Current status, challenges, and opportunities,'' \emph{Renew. Sust.
  Energ. Rev.}, vol. 143, p. 110909, Jun. 2021.

\bibitem{4-583}
Z.~Li, S.~Jazebi, and F.~De~Leon, ``Determination of the optimal switching
  frequency for distribution system reconfiguration,'' \emph{IEEE Trans. Power
  Del.}, vol.~32, no.~4, pp. 2060--2069, Aug. 2016.

\bibitem{4-905}
C.~Wang, S.~Lei, P.~Ju, C.~Chen, C.~Peng, and Y.~Hou, ``{MDP}-based
  distribution network reconfiguration with renewable distributed generation:
  An approximate dynamic programming approach,'' \emph{IEEE Trans. Smart Grid},
  vol.~11, no.~4, pp. 3620--3631, Jul. 2020.

\bibitem{4-1302}
N.~Hatziargyriou and \textit{et~al.}, ``Definition and classification of power
  system stability {\textendash} revisited {\&} extended,'' \emph{IEEE Trans.
  Power Syst.}, vol.~36, no.~4, pp. 3271--3281, Jul. 2021.

\bibitem{4-990}
M.~Farrokhabadi and \textit{et~al.}, ``Microgrid stability definitions,
  analysis, and examples,'' \emph{IEEE Trans. Power Syst.}, vol.~35, no.~1, pp.
  13--29, Jan. 2020.

\bibitem{4-891}
P.~Vorobev, P.-H. Huang, M.~A. Hosani, J.~L. Kirtley, and K.~Turitsyn,
  ``High-fidelity model order reduction for microgrids stability assessment,''
  \emph{IEEE Trans. Power Syst.}, vol.~33, no.~1, pp. 874--887, Jan. 2018.

\bibitem{4-620}
D.~K. Dheer, O.~V. Kulkarni, S.~Doolla, and A.~K. Rathore, ``Effect of
  reconfiguration and meshed networks on the small-signal stability margin of
  droop-based islanded microgrids,'' \emph{IEEE Trans. Ind. Appl.}, vol.~54,
  no.~3, pp. 2821--2833, May 2018.

\bibitem{4-783}
M.~Khanabadi, Y.~Fu, and C.~Liu, ``Decentralized transmission line switching
  for congestion management of interconnected power systems,'' \emph{IEEE
  Trans. Power Syst.}, vol.~33, no.~6, pp. 5902--5912, Nov. 2018.

\bibitem{4-822}
F.~Ding and K.~A. Loparo, ``Feeder reconfiguration for unbalanced distribution
  systems with distributed generation: A hierarchical decentralized approach,''
  \emph{IEEE Trans. Power Syst.}, vol.~31, no.~2, pp. 1633--1642, Mar. 2016.

\bibitem{4-995-ea}
T.~Han, Y.~Song, and D.~J. Hill, ``Ensuring network connectedness in optimal
  transmission switching problems,'' \emph{{IEEE} Trans. Circuits Syst. {II}},
  vol.~68, no.~7, pp. 2603--2607, Jul. 2021.

\bibitem{4-1163}
C.~Coffrin and P.~V. Hentenryck, ``A linear-programming approximation of {AC}
  power flows,'' \emph{{INFORMS} Journal on Computing}, vol.~26, no.~4, pp.
  718--734, Nov. 2014.

\bibitem{4-1129}
W.~E. Brown and E.~Moreno-Centeno, ``Transmission-line switching for load shed
  prevention via an accelerated linear programming approximation of {AC} power
  flows,'' \emph{IEEE Trans. Power Syst.}, vol.~35, no.~4, pp. 2575--2585, Jul.
  2020.

\bibitem{4-845}
M.~Thomas, \emph{Power system SCADA and smart grids}.\hskip 1em plus 0.5em
  minus 0.4em\relax Boca Raton, FL: CRC Press, 2015.

\bibitem{4-761}
J.~Ostrowski, J.~Wang, and C.~Liu, ``Transmission switching with
  connectivity-ensuring constraints,'' \emph{IEEE Trans. Power Syst.}, vol.~29,
  no.~6, pp. 2621--2627, Apr. 2014.

\bibitem{4-1283}
P.~Li, X.~Huang, J.~Qi, H.~Wei, and X.~Bai, ``A connectivity constrained {MILP}
  model for optimal transmission switching,'' \emph{IEEE Trans. Power Syst.},
  vol.~36, no.~5, pp. 4820--4823, Sep. 2021.

\bibitem{4-1003}
C.~Liu, J.~Wang, and J.~Ostrowski, ``Static switching security in multi-period
  transmission switching,'' \emph{IEEE Trans. Power Syst.}, vol.~27, no.~4, pp.
  1850--1858, Nov. 2012.

\bibitem{4-1365}
S.~Santos, M.~Gough, D.~Z. Fitiwi, J.~Pogeira, M.~Shafie-khah, and J.~Catalao,
  ``Dynamic distribution system reconfiguration considering distributed
  renewable energy sources and energy storage systems,'' \emph{IEEE Syst. J.},
  pp. 1--11, Jan. 2022.

\bibitem{4-848}
T.~McDermott, I.~Drezga, and R.~Broadwater, ``A heuristic nonlinear
  constructive method for distribution system reconfiguration,'' \emph{IEEE
  Trans. Power Syst.}, vol.~14, no.~2, pp. 478--483, May 1999.

\bibitem{4-847}
F.~Gomes, S.~Carneiro, J.~Pereira, M.~Vinagre, P.~Garcia, and L.~Araujo, ``A
  new heuristic reconfiguration algorithm for large distribution systems,''
  \emph{IEEE Trans. Power Syst.}, vol.~20, no.~3, pp. 1373--1378, Aug. 2005.

\bibitem{4-838}
S.~Goswami and S.~Basu, ``A new algorithm for the reconfiguration of
  distribution feeders for loss minimization,'' \emph{IEEE Trans. Power
  Deliv.}, vol.~7, no.~3, pp. 1484--1491, Jul. 1992.

\bibitem{4-1002}
\BIBentryALTinterwordspacing
T.~Han, ``Structure-oriented optimization and control,'' 2021. [Online].
  Available: \url{https://github.com/thanever/SOC/tree/master/Ott/Data}
\BIBentrySTDinterwordspacing

\bibitem{4-1000}
D.~Lee, K.~Turitsyn, D.~K. Molzahn, and L.~A. Roald, ``Feasible path
  identification in optimal power flow with sequential convex restriction,''
  \emph{IEEE Trans. Power Syst.}, vol.~35, no.~5, pp. 3648--3659, Sep. 2020.

\end{thebibliography}
